\documentclass[final,3p,times,twocolumn]{elsarticle}
\usepackage{graphicx}
\usepackage{amssymb}
\usepackage[utf8]{inputenc}
\usepackage{textcomp}

\usepackage{booktabs}
\usepackage{rotating}
\usepackage{multirow}
\usepackage{caption}
\usepackage{subcaption}

\usepackage{amsmath}

\usepackage{url,hyperref}

\usepackage{color}
\usepackage[usenames,dvipsnames,svgnames,table]{xcolor}

\journal{Journal of Nuclear Materials }

\begin{document}

\begin{frontmatter}



\title{Sink strength calculations of dislocations and loops using OKMC}


\author[sck,helsinki]{V.~Jansson\corref{cor1}}
\ead{ville.b.c.jansson@gmail.com}
\author[sck]{L.~Malerba}
\author[lille]{A. de Backer}
\author[lille,edfcnrs]{C.S. Becquart}
\author[edf,edfcnrs]{C. Domain}

\cortext[cor1]{Corresponding author. Tel. +32 1433 3096, fax: +32 1432 1216.}

\address[sck]{Institute of Nuclear Materials Science, SCK$\bullet$CEN, Boeretang
200, 2400
{\sc Mol, Belgium}}
\address[helsinki]{Department of Physics, P.O. Box 43 (Pehr
Kalms gata 2), FI-00014 {\sc University of Helsinki, Finland}}
\address[lille]{Unit\'e Mat\'eriaux Et Transformations (UMET), UMR CNRS 8207, Universit\'e de Lille 1, ENSCL, {\sc F-59655 Villeneuve d'Ascq Cedex, France}}
\address[edf]{EDF-R\&D, D\'epartement Mat\'eriaux et M\'ecanique des Composants (MMC), Les Renardi\'eres, {\sc F-77818 Moret sur Loing Cedex, France}}
\address[edfcnrs]{Laboratoire commun EDF-CNRS Etude et Mod\'elisation des Microstructures pour le Vieillissement des Mat\'eriaux (EM2VM), {\sc France}}

\begin{abstract}

We calculate the sink strength of dislocations and toroidal absorbers using Object Kinetic Monte Carlo and compare with the theoretical expressions. We get good agreement for dislocations and loop-shaped absorbers of 3D migrating defects, provided that the volume fraction is low, and fair agreements for dislocations with 1D migrating defects. The master curve for the 3D to 1D transition is well reproduced with loop-shaped absorbers and fairly well with dislocations. We conclude that, on the one hand, the master curve is correct for a wide range of sinks and that, on the other, OKMC techniques inherently take correctly into account the strengths of sinks of any shape, provided that an effective way of appropriately inserting the sinks to be studied can be found. 
\end{abstract}


\end{frontmatter}

\frenchspacing

% \linenumbers

\section{Introduction}\label{sec:intruction}

% Motivation: material and defects
Irradiation introduces mobile defects such as self-interstitial atom (SIA) clusters and vacancy clusters in metals. These defects will interact with each other and with the pre-existing microstructure, mainly dislocations, thereby inducing nanostructural changes that will affect the mechanical properties of the material. To fully understand the evolution of the defect populations over time, the rates of these reactions needs to be correctly assessed.

% OKMC
Object Kinetic Monte Carlo (OKMC) is a stochastic simulation method, where the dynamic behaviour of all objects, such as SIA or vacancy clusters, is described by pre-defined probabilities. It is a well-suited technique for simulation of the evolutions of radiation induced defects in iron alloys (See \textit{e.g.} \cite{jansson2013simulation}). The OKMC has been shown to be equivalent to rate theory calculations \cite{stoller2008mean}. It has the advantage of going beyond the mean-field approximation and taking explicitly all spatial correlations, except the elastic interactions, into account. 

% The importance of k^2, What does it depend on?
In mean-field approaches, the rate at which a mobile defect interacts with a cluster or dislocation of a given shape and size, acting as a sink, is given by the sink strength, $k^2$. This is proportional to the inverse square of the average distance covered by the defects before the interaction, which is normally absorption or clustering. The sink strength is \textit{a priori} affected by the shape and size of the sinks, their number density, their type and orientation. Also, the migration regime of the defects will have an impact: defects that migrate in a 1D fashion are less likely to interact with sinks than defects that migrate in 3D or in a fashion between fully 1D and 3D.

Analytical expressions for different sink shapes, such as spheres, toroids and dislocations have been theoretically obtained in the case of 3D migrating defects and a number of them is reviewed by \textit{e.g.} F.A. Nichols in \cite{nichols1978estimation}. Barashev \textit{et al}. derived expressions for fully 1D migrating defects in the case of spherical absorbers, dislocations and grain boundaries \cite{barashev2001reaction}. The 3D to 1D transition has been studied by Trinkaus \textit{et. al.}, who also proposed a master curve for the transition \cite{trinkaus20021d,trinkaus2004reaction}. Malerba \textit{et. al.} \cite{malerba2007object} showed that OKMC calculations of the sink strength for spherical sinks and grain boundaries show good agreement with analytical expressions used in rate theory and that also the 3D to 1D defect migration regime transition with spherical absorbers can be reproduced using OKMC. These results simultaneously corroborate the theory and show the equivalence between OKMC and 
mean-field approaches.

% Research statement
In this work, we extend the study by Malerba \textit{et. al.} \cite{malerba2007object}, where spherical absorbers were considered, to also calculate the sink strength for dislocation lines and toroidal absorbers, the latter corresponding to dislocation loops, and compare the results with analytical expressions available from rate theory. The structure of this paper is as follows: In section \ref{sec:methods}, we describe our methodology. In Sec. \ref{sec:dislocations} we compare the sink strength of dislocations obtained by OKMC to rate theory expressions, in the limits of 3D and 1D migration, and in Sec. \ref{sec:loops}, we do the same with toroidal absorbers. Finally, in Sec. \ref{sec:transition}, we study the transition of the migration regime from 3D to 1D of defects absorbed by dislocations and toroidal sinks and compare with the theoretical master curve. The discussion and conclusions are found in Sec. \ref{sec:discussion} and \ref{sec:conclusions}, respectively.

\section{Computation method}\label{sec:methods}

We have estimated the sink strength of straight dislocations and dislocation loops using the same methods as in \cite{malerba2007object} (where spherical absorbers were considered), which is here briefly recalled. For our calculation we use the OKMC code LAKIMOCA \cite{domain2004simulation}. The probabilities for migration jumps of defects in the simulations are given in terms of Arrhenius frequencies for thermally activated events, $\Gamma_i = \nu_i\exp \left(\frac{-E_{a,i}}{k_B T}\right)$, where $\nu_i$ is the attempt frequency, $E_{a,i}$ the activation energy for the process, $k_B$ Boltzmann's constant and $T$ the temperature. Events are randomly chosen according to their probability, following the Monte Carlo algorithm \cite{metropolis1953equation}. The simulated time is increased according to the resident time algorithm \cite{young1966monte} with $\Delta t = 1/\left(\sum_{i=1}^{N_{int}} \Gamma_i + \sum_{j=1}^{N_{ext}}P_j\right)$, where $N_{int}$ is the number of internal events such as defect jumps and 
$N_{ext}$ the number of external events, such as cascades or Frenkel pair creation, with $P_j$ being the probabilities for the external events. In the long term, this equation substitutes $\Delta t' = -\ln{u}\Delta t$ , which is fully exact by including the stochasticity due to the Poisson distribution \cite{bortz1975new}. $u$ is here a uniform random number between 1 and 0 . 

Objects, such as defects or clusters, in the model are described as geometrical objects, such as spheres, toroids or cylinders. Reactions between objects take place when these overlap geometrically. Reactions could be annihilations or clustering. In this study, the only reaction considered is absorption, where the reacting defect will not change the volume of the absorbing objects.

The straight dislocation was simulated as an immobile cylinder-shaped sink whose two opposite faces touch the faces of a non-cubic simulation box. No defect can ever impinge on the cylinder from one of the faces, as no defect is allowed to be outside the box, so this is an effective way to simulate an infinitely long straight dislocation. However, since periodic boundary conditions are applied in all direction, we are in practice simulating a regular array of infinitely long straight dislocations: this must be taken into account to rationalize the results and certainly when choosing the theoretical expression to which the simulation results are to be compared. (On the other hand, as discussed by Brailsford and Bullough \cite{brailsford1981theory}, there is no real theoretical expression for the sink strength of an array of dislocations significantly different from a lattice and for real random arrays the $Z$ factor of proportionality with the dislocation density should be regarded as an empirical parameter.)

\begin{figure}
 \centering
  \includegraphics[width=\columnwidth]{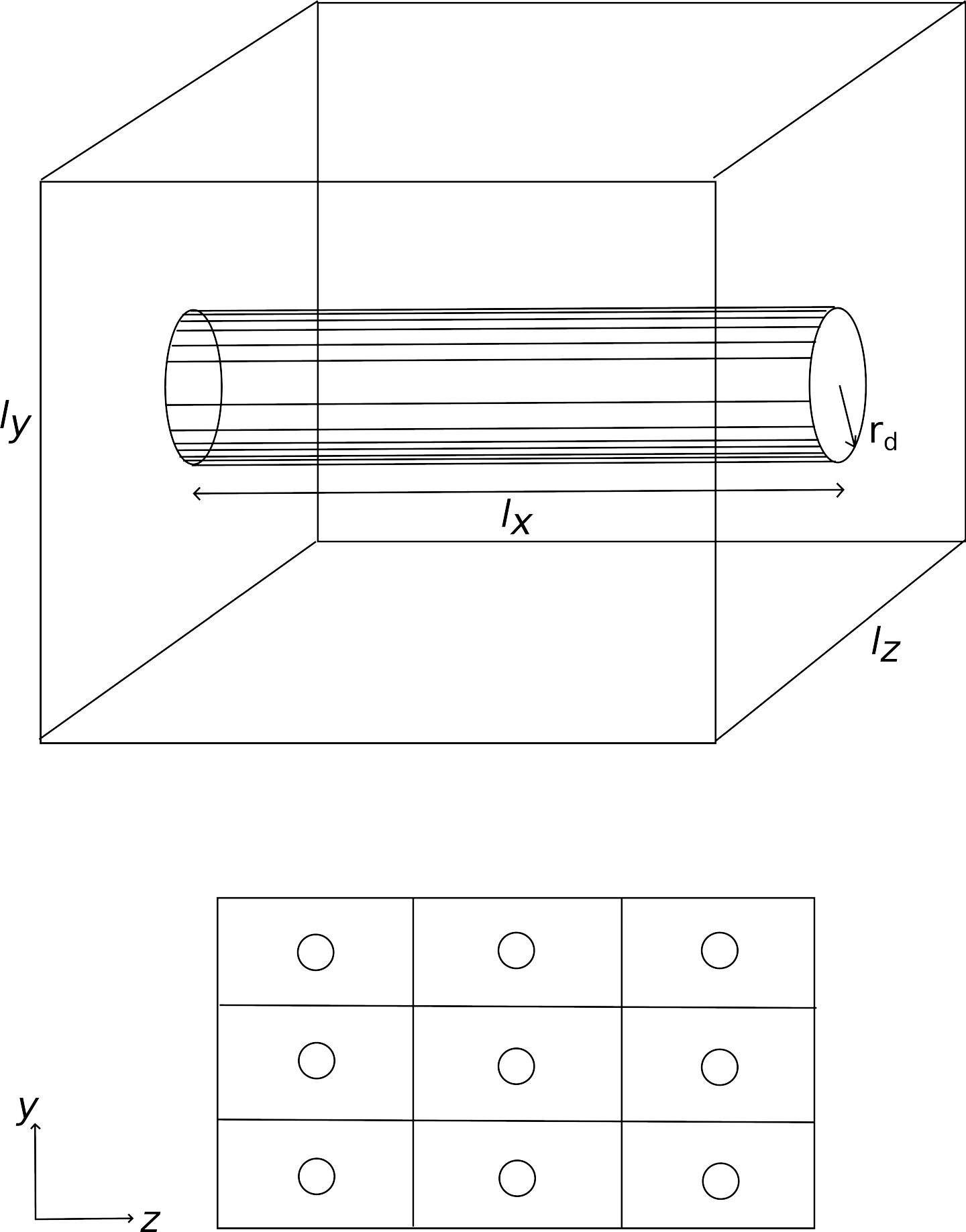} % 1.jpg
  \caption{Upper side: Pictorial representation of the non-cubic simulation box and the cylinder mimicking the sink effect of a straight dislocation. Lower side: view normal to $x$ axis of the simulation box and image boxes corresponding to applying periodic boundary conditions: the image dislocations effectively create a regular array.}
  \label{1.jpg}
\end{figure}

The simulation box with the dislocation is pictorially represented in Fig \ref{1.jpg}. Different dislocation densities and capture radii, $r_d$, (as in Fig \ref{1.jpg}) were explored. With reference to Fig. \ref{1.jpg}, the dislocation density, $\rho_d$, was changed by varying $l_y$ and  $l_z$: $\rho_d = (l_y l_z)^{-1}$, for a fixed  $l_x=100a_0$ (since iron is taken as materials of reference, the underlying lattice is body-centred-cubic (bcc) and the lattice parameter is $a_0=2.87$ Å), with $l_x \neq l_y\neq l_z$. Dislocation densities were varied between $10^{14}$ and $10^{15}$ m$^{-2}$, capture radii, $r_d$, between 2 and 9 nm (smaller densities, down to $5\cdot10^{13}$ m$^{-2}$, were also considered in the 3D limit only; simulations with smaller radii, 0.5 and 1 nm, for $\rho_d=10^{14}$ m$^{-2}$ did not provide sufficient statistics to be fully acceptable in the 1D limit). The simulation temperature was arbitrarily set to 573 K, but it does not have any influence on the sink strength calculation. .

\begin{figure}
 \centering  
 \includegraphics[width=\columnwidth]{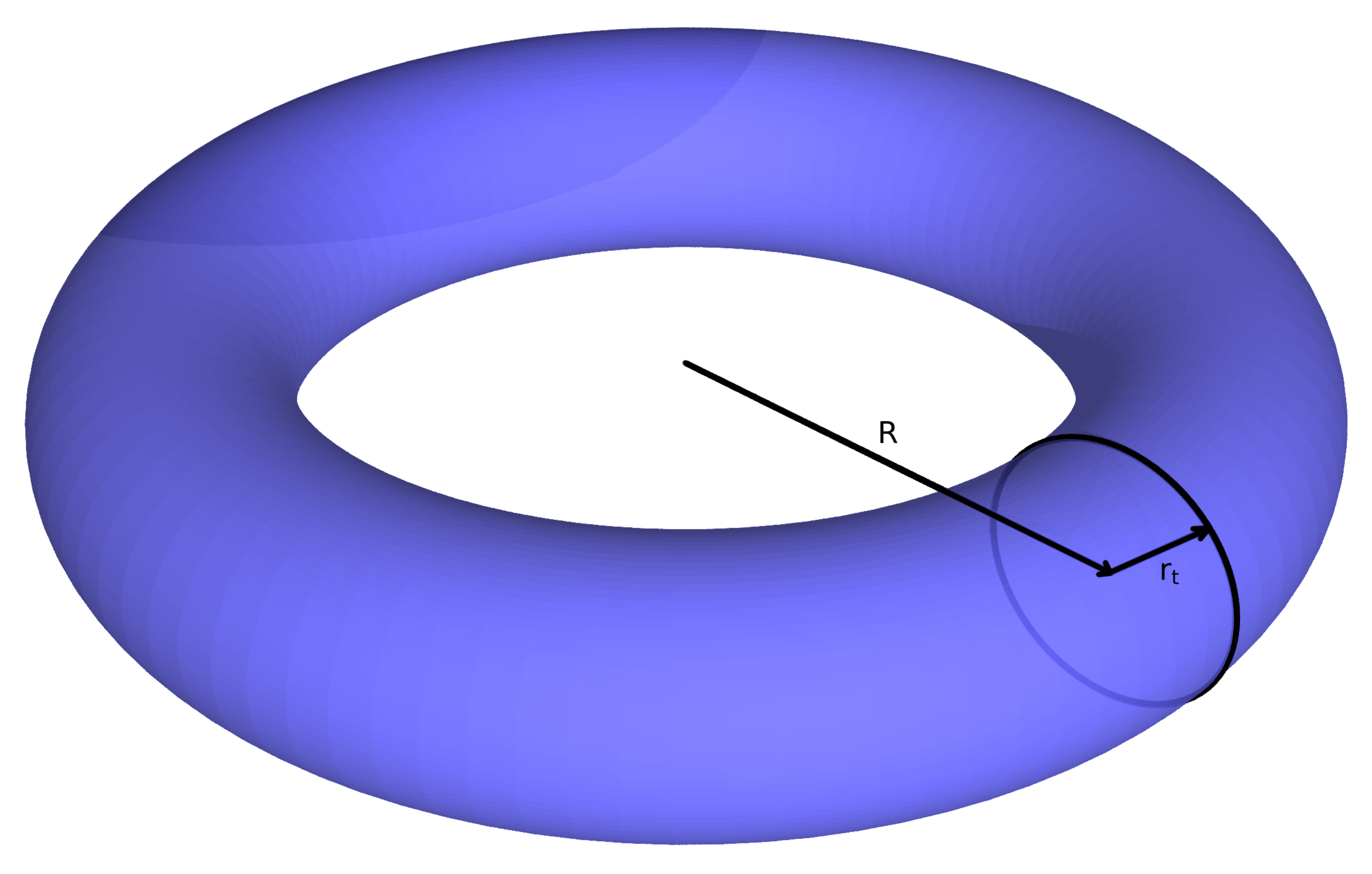}
  \caption{Depiction of a toroid with the major radius $R$ and the minor radius $r_t$.}
  \label{fig:torus}
\end{figure}

Dislocation loops are simulated by immobile sinks of toroidal shape, depicted in Fig. \ref{fig:torus}, and characterized by the major radius, $R$, the minor radius $r_t$ and their orientation with four possible [111] Burgers vectors. Toroidal sinks with the same $R$ and $r$ were randomly distributed in the simulation box with random orientations for a given number density. The simulation box size used for the loop sink strength calculations were $300\times350\times400a_0^3$, except in Sec. \ref{sec:orientation} and Sec. \ref{sec:transition}, were smaller box sizes of $150\times200\times250a_0^3$ and $250\times300\times350a_0^3$, respectively, were used in the loop cases. A simulation temperature of 373 K was arbitrarily used for all loop cases.

The sink strength is obtained as \cite{malerba2007object}
\begin{equation}\label{eq:sink_strength}
 k^2 = \frac{2n}{d^2_j\langle n_j\rangle},
\end{equation}
where $\langle n_j \rangle$ is the average number of jumps performed by the defect, introduced one by one in the simulation. $n$ is the dimensionality of the motion and $d_j$ is the jump distance, defined in the bcc lattice as the first nearest neighbour distance,
\begin{equation}
 d_j = \frac{\sqrt{3}}{2}a_0.
\end{equation}
In the simulations, 3D to 1D migrating defects were introduced at a random position one by one. 
The sink strength is then calculated using the average number of jumps needed to be absorbed by a sink, $\langle n_j\rangle$, and Eq. \eqref{eq:sink_strength}. The defects have a spherical capture radius of $r_0=2.5$ Å. The attempt frequency used is $8.071\cdot10^{13}$ s$^{-1}$ and the migration energy 0.31 eV, corresponding to a single SIA \cite{jansson2013simulation,anento2010atomistic,takaki1983resistivity}. Clustering of defects was explicitly forbidden. 

For the dislocation case, the simulation was stopped when one of the two following conditions was fulfilled: (a) 30000 defects had been introduced and disappeared at the sinks (in practice, somewhat less than this, because the defects that happened to be created inside the sink are disregarded); (b) the sum of the number of jumps taken by all the defects introduced in the box equals $10^{15}$. Generally the first condition dictated the end of the simulation, but for small volume fractions in the 1D case sometimes the second one overruled. For the loop case, the simulation was stopped when 10000 defects had been introduced and absorbed by the sinks.

The 3D to 1D regime transition was explored by increasing the number of 1D jumps after which the defect changes direction from 1 (effective 3D migration) to $10^{10}$--$10^{11}$ (effective 1D migration). In addition for the straight dislocation case, the 1D limit was also simulated by requiring an energy of change of direction of 2 eV at 573 K, which corresponds in practice to no change of direction in the course of the whole simulation, at the temperatures considered (the effectiveness of this way of operating to reach the 1D limit, using a non-cubic box, is demonstrated and discussed in \cite{malerba2007object}).

\section{Results}\label{sec:results}

\subsection{Straight dislocations}\label{sec:dislocations}

\subsubsection{3D limit for dislocations}

\begin{figure}
 \centering  
 \includegraphics[width=\columnwidth]{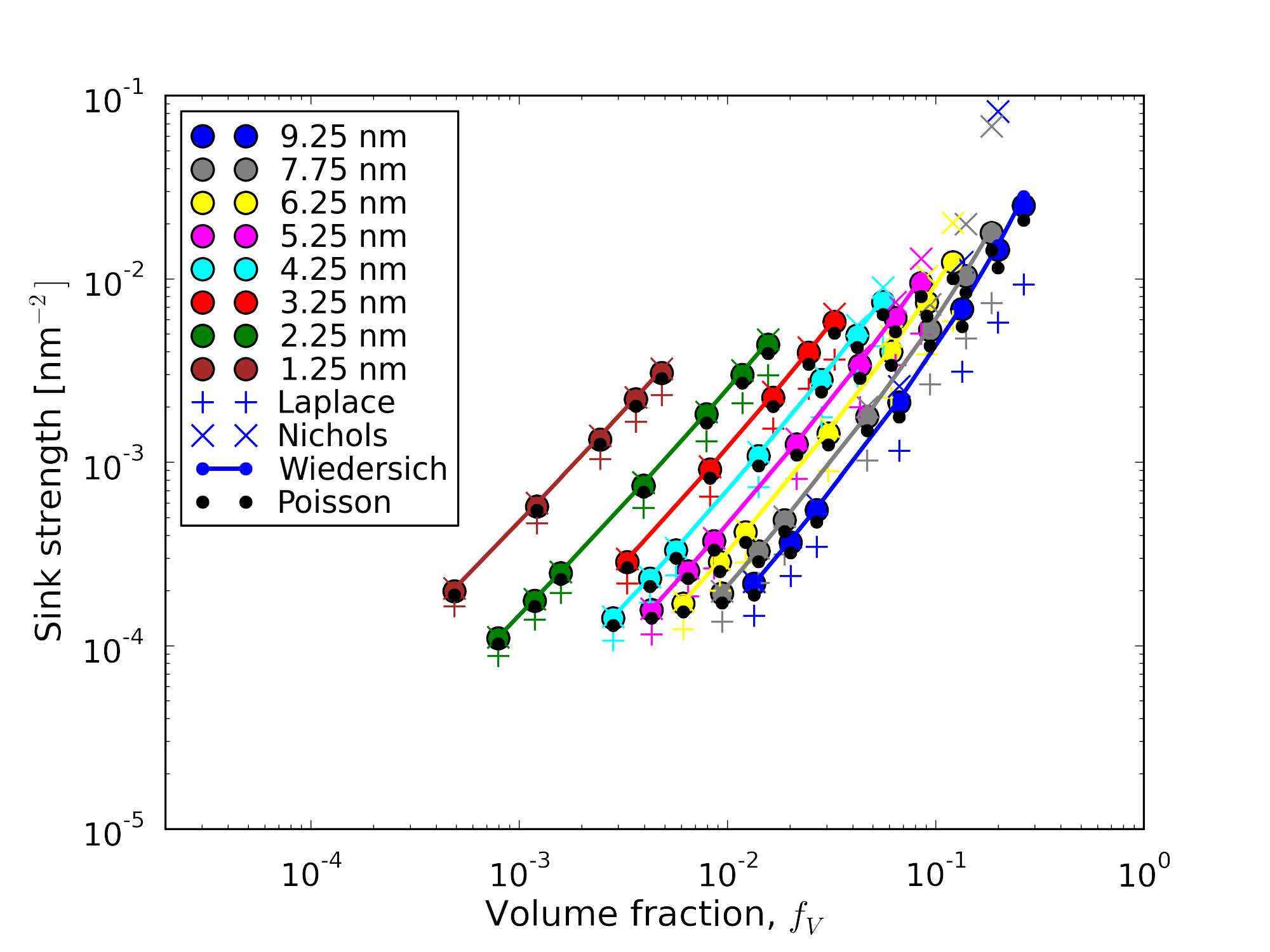}
  \caption{OKMC dislocation sink strengths in the 3D limit versus volume fraction. The capture radius, $r_d$, is varied between 1.25 and 9.25 nm. The dislocation densities are specified in the text. Points from different theoretical expressions are superposed with the corresponding colours. Only the expression of Wiedersich (\textit{Cf.} Table \ref{table:dislocation_eq}) matches all simulation points.}
  \label{2.jpg}
\end{figure}

Fig. \ref{2.jpg} shows the ``cloud'' of sink strength points hitherto obtained from the simulation in the 3D limit, in the above-specified ranges of capture radii and dislocation densities, plotted versus volume fraction. Patterns according to which the data points can be grouped are clearly recognisable: they correspond to the same capture radius (indicated by the same colour) or the same dislocation density. On the same figure the points calculated for the same dislocation density and capture radius using different approximations, as found in \cite{nichols1978estimation} (\textit{Cf.} Table \ref{table:dislocation_eq}), are indicated. All the theoretical expressions are for a regular array of parallel dislocations. It can be seen that, while for small volume fractions all expressions are acceptably valid, for large volume fractions only the expression deduced by Wiedersich (\textit{Cf.} Table \ref{table:dislocation_eq})
remains valid and in agreement with the simulation results. The expressions obtained by solving either Laplace or Poisson equations underestimate the sink strength, while the approximated expression proposed by Nichols in \cite{nichols1978estimation} (\textit{Cf.} Table \ref{table:dislocation_eq}) exhibits a singularity in the range of volume fractions of interest and starts diverging above a certain volume fraction.
\begin{table}
\centering
\caption{Analytical expressions for the sink strength of dislocation lines. Here $\rho = r_d\sqrt{\pi\rho_d}$ \cite{nichols1978estimation}.}
\label{table:dislocation_eq}
\begin{tabular*}{\columnwidth}{@{\extracolsep{\fill}} l c p{0.7\columnwidth}}
 \toprule
 & $k^2_{d,3}$\\
 \midrule
 \vspace{0.5cm} Wiedersich \cite{wiedersich1972theory} &
$\frac{2\pi\rho_d(1-\rho^2)}{\ln\left(\frac{1}{\rho}\right) - \frac{3}{4} + \frac{1}{4}\rho^2(4-\rho^2)}$\\
\vspace{0.5cm} Laplace &
$\frac{2\pi\rho_d}{\ln\left(\frac{1}{\rho}\right)}$\\
\vspace{0.5cm} Poisson &
$\frac{2\pi\rho_d}{\ln\left(\frac{1}{\rho}\right)-\frac{1}{2}+\frac{1}{2}\rho^2}$\\
Nichols &
$\frac{2\pi\rho_d}{\ln\left(\frac{1}{\rho}\right)-\frac{3}{4}}$\\
\bottomrule
\end{tabular*}
\end{table}

\begin{figure} 
 \centering
  \includegraphics[width=\columnwidth]{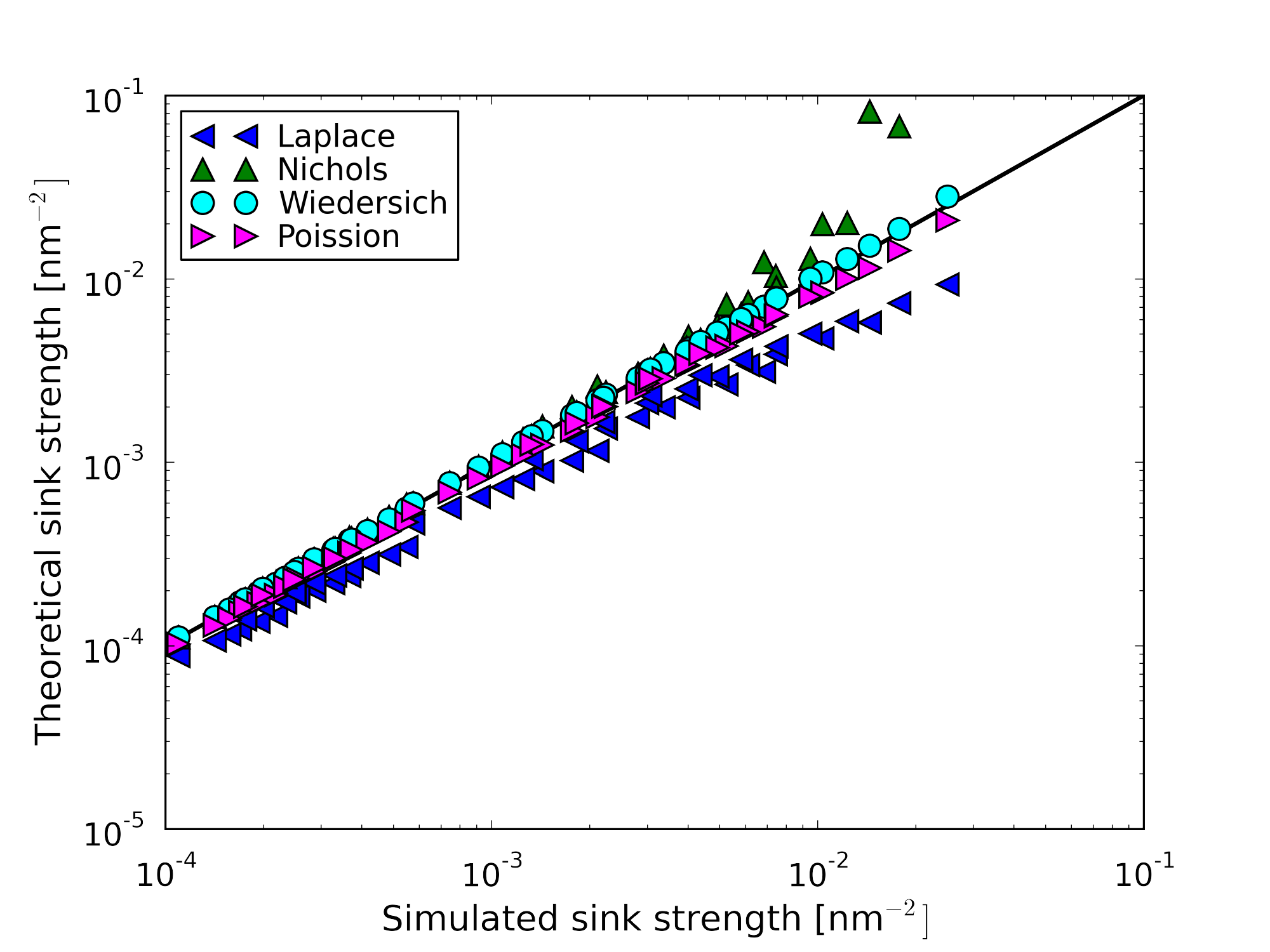}
  \caption{Comparison between simulation values and different theoretical values for the sink strength of a regular array of parallel dislocations in the 3D limit.}
  \label{3.jpg}
\end{figure}

In Fig. \ref{3.jpg} the comparison between OKMC simulation and different theoretical predictions is made directly on a 45\textdegree{} straight line. As can be seen, in the case of the Wiedersich expression (\textit{Cf.} Table \ref{table:dislocation_eq}), the points fall perfectly on the line, so we can state that there is perfect correspondence between simulation and theory.


\subsubsection{1D limit for dislocations}

\begin{figure}
 \centering
  \includegraphics[width=\columnwidth]{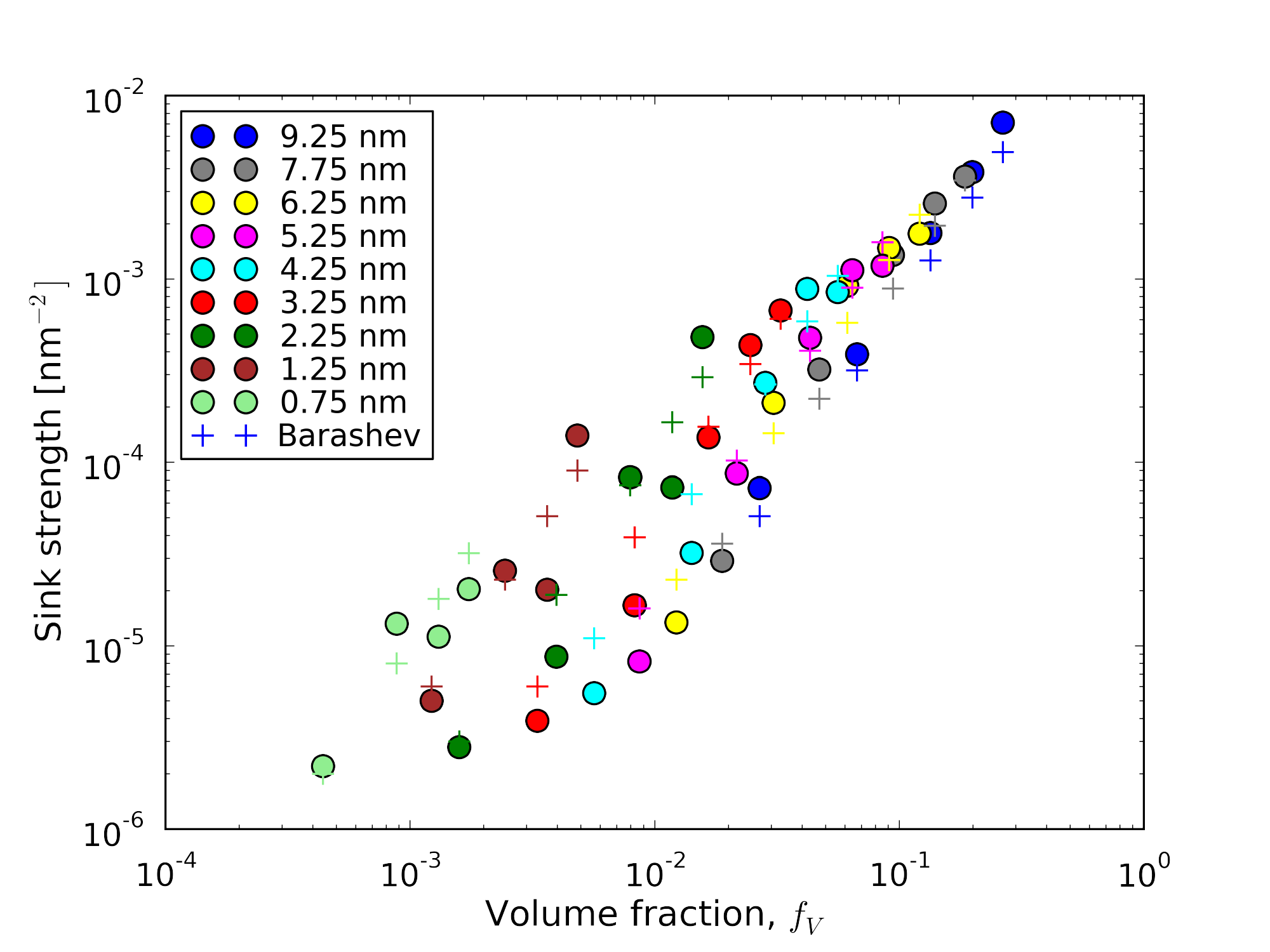}
  \caption{OKMC dislocation sink strength in the 1D limit versus volume fraction in the range of capture radii, $r_d$, and dislocation densities specified in the text, compared with the theoretical points obtained from Eq. \eqref{eq:barashev}.}
  \label{4.jpg}
\end{figure}

\begin{figure}
 \centering
  \includegraphics[width=\columnwidth]{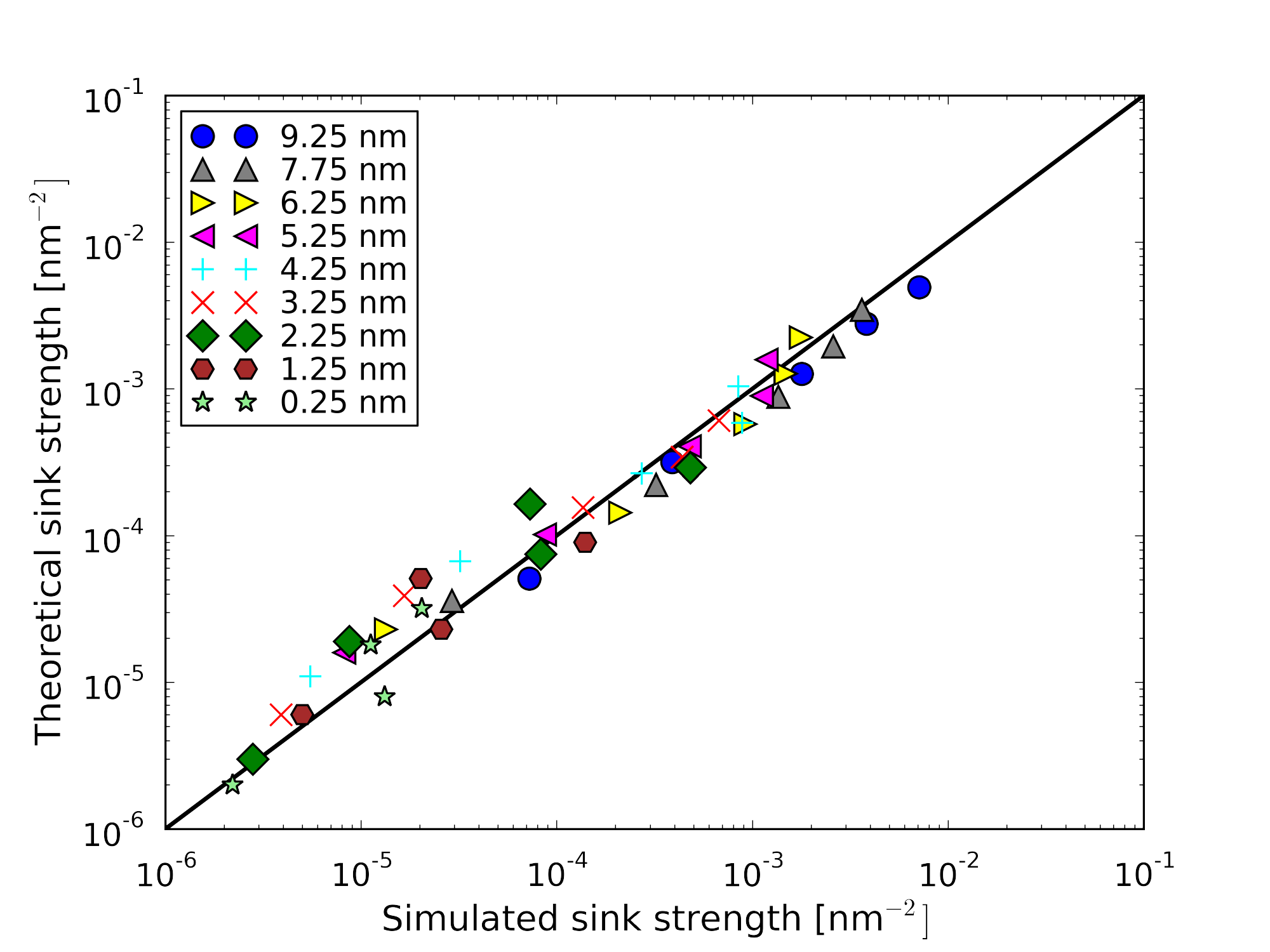}
  \caption{Comparison between theory, Eq. \eqref{eq:barashev}, and simulation values for the sink strength of a regular array of parallel dislocations in the 1D limit. The colours correspond to different capture radii, $r_d$ = 0.75--9.26 nm, and are the same as in Fig. \ref{4.jpg}.}
  \label{5.jpg}
\end{figure}

Figs \ref{4.jpg} and \ref{5.jpg} are the equivalent to Figs. \ref{2.jpg} and \ref{3.jpg}, comparing the simulation results with theory using the ``cloud'' and the 45\textdegree{} line representation, respectively, for 1D migrating defects. The only theoretical expression used in this case for comparison is the one proposed by A. Barashev \textit{et al.} in \cite{barashev2001reaction}:
\begin{equation}\label{eq:barashev}
 k_{d,1}^2 = 3\cdot2(\pi r_d \rho^\ast)^2,
\end{equation}
where the factor 3 (which does not appear in the mentioned reference) comes from the fact of using here the 3D diffusion coefficient (in order to consistently trace the transition between regimes) and $\rho^\ast$ is defined as ``the mean number of dislocations lines intersecting a unit area (surface density)''. In the present case of regular array of parallel dislocations, $\rho^\ast=\rho_d$: there is indeed only one surface crossed by dislocations and the surface density is (with reference to Fig. \ref{1.jpg}) $\rho^\ast = (l_y l_z)^{-1}$, which is coincident with the dislocation density. This equality has been therefore used to apply Eq.. \eqref{eq:barashev} and to compare it with the simulation results in the figures. It can be seen that the comparison is acceptable, particularly 
in Fig. \ref{5.jpg}, although it is certainly not as good as in the 3D case. At the same time, in Fig. \ref{4.jpg} the patterns according to which the data points should be grouped by capture radii or dislocation densities are less easy to spot, a sign of larger scatter. We shall address later on the problem of establishing where this scatter and the less good agreement between theory and simulation in this 1D case may come from.

\subsection{Loops}\label{sec:loops}

If the rate of annihilation at sinks is defined as $Dck^2$, where $D$ is the 3D-diffusion coefficients, $c$ the defect concentration and $k^2$ the sink strength, the theoretical expression of sink strength for toroidal absorbers can be derived from the work of F.A. Nichols \cite{nichols1978estimation} as:
\begin{equation}\label{eq:theory_torids}
 k^2_{t,3} = \frac{4 \pi^2 n (R^2 -r_t^2)^{1/2} }{ \ln(8 R / r_t) },
\end{equation}
where $R$ is the major radius and $r_t$ the minor radius, as shown in Fig. \ref{fig:torus}, and $n$ is the number density of toroidal sinks: $R \gtrsim 3r_t$ is the condition for the applicability of Eq. \eqref{eq:theory_torids}.
For comparison, the sink strength of spherical absorbers with a radius $r_s$ and a density $n$ for 3D migrating defects is given by \cite{malerba2007object,brailsford1981theory,brailsford1979effect}:
\begin{equation}\label{eq:theory_spheres}
 k^2_{s,3} = 4\pi n r_s \left(1+ r_s\sqrt{4\pi n r_s}\right). 
\end{equation}
For 1D migrating defects, the sink strength is given by
\begin{equation}\label{eq:theory_spheres_1d}
 k^2_{s,1} = 6(\pi r_s^2n)^2.
\end{equation}

In this section we will calculate the sink strength for a population of different number densities, $n$, different $r$ and $R$ values and compare with the theory for toroidal and spherical sinks. 

\subsubsection{The dependence on the sink number density}

\begin{figure}
 \centering
  \includegraphics[width=\columnwidth]{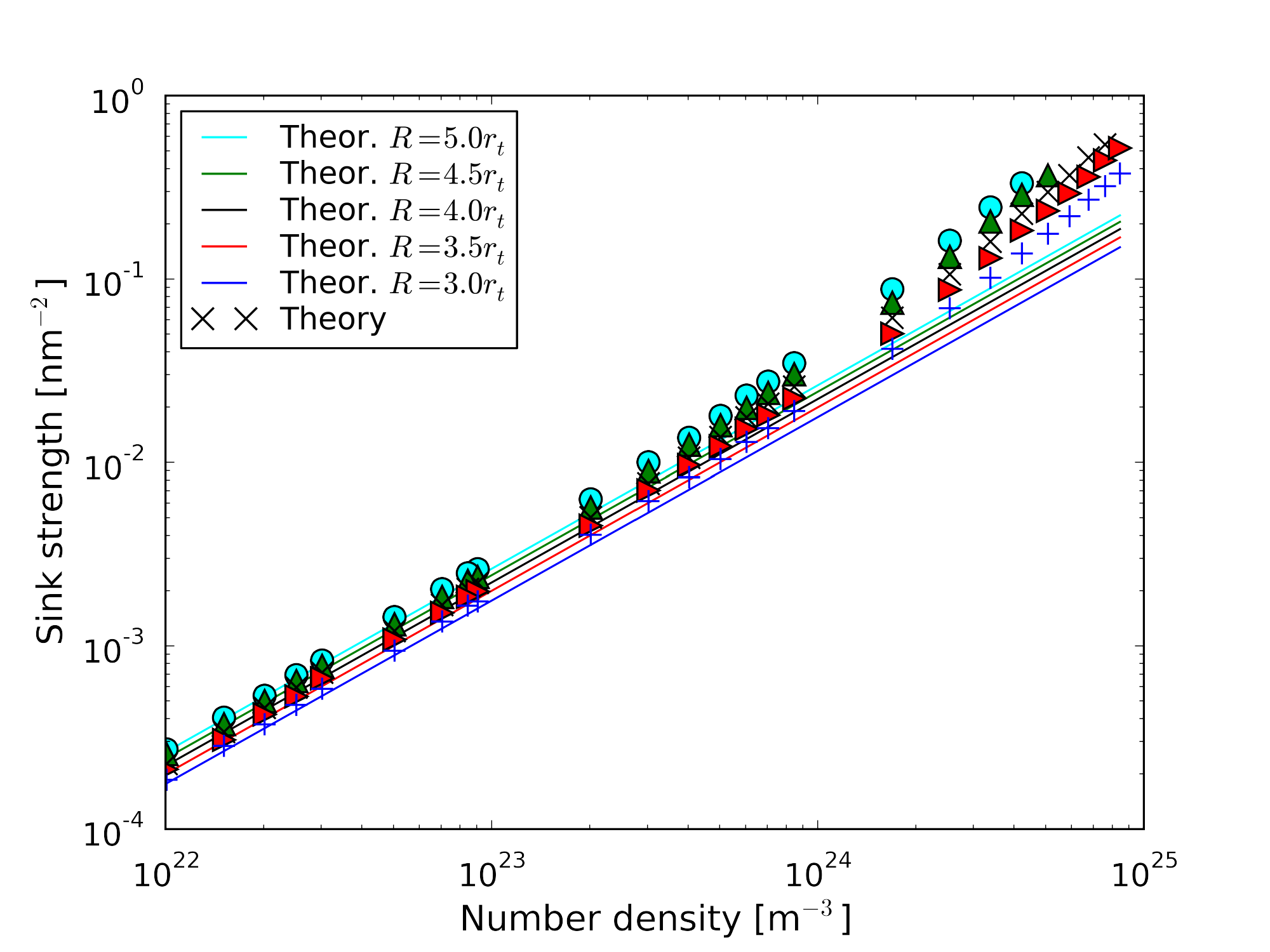}
  \caption{Sink strength of loops for different ratios between radii when the major one increases, as a function of number density: both OKMC calculations and theoretical expressions collapse onto essentially the same curve.}
  \label{fig:P20100824-1.pdf}
\end{figure}

\begin{figure}
 \centering
  \includegraphics[width=\columnwidth]{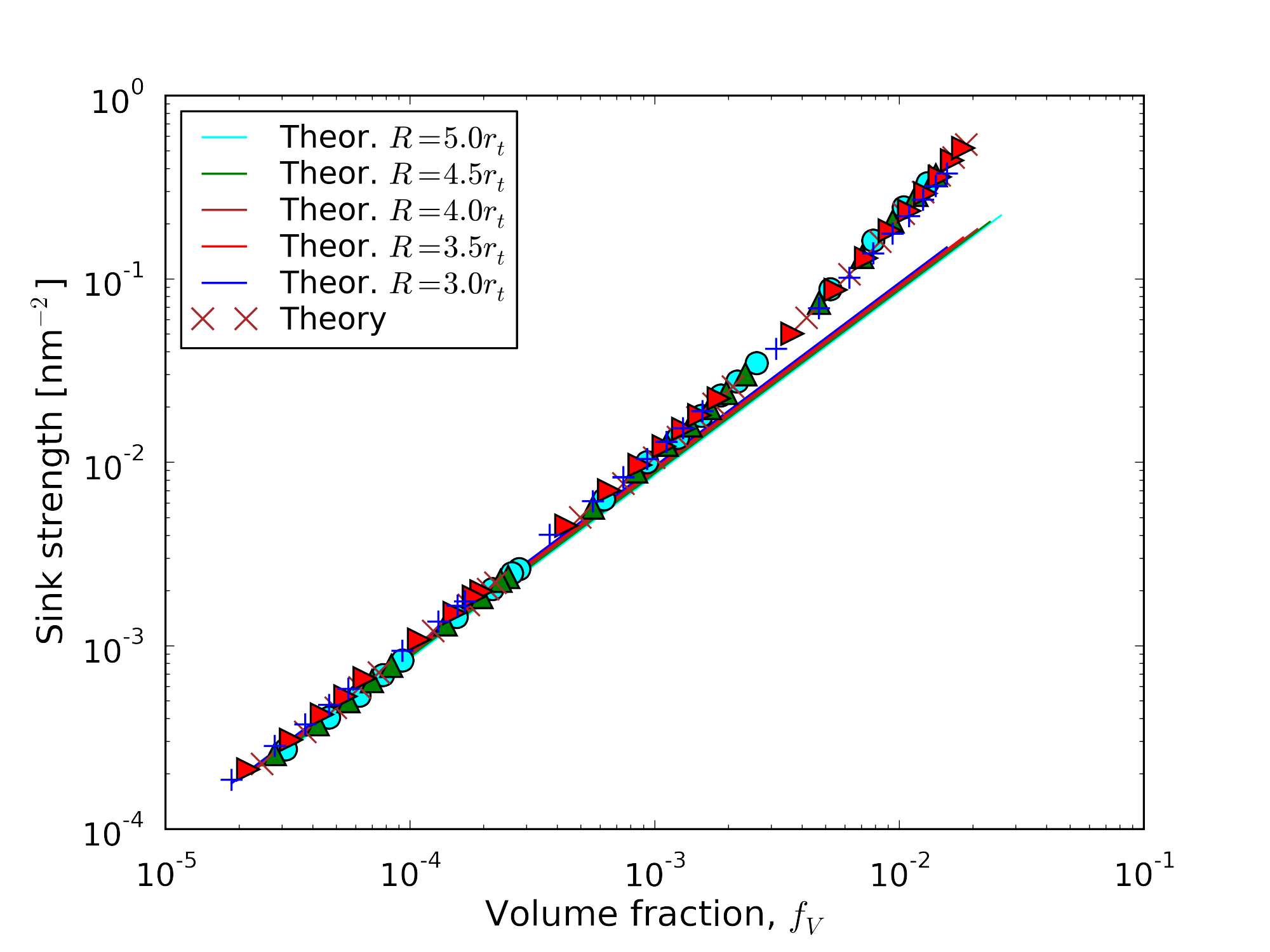}
  \caption{The effect of the loop number density in terms of volume fraction and compared to the theory for toroidal sinks}
  \label{fig:P20100824-1_Vfrac.pdf}
\end{figure}

We fix the minor radius as $r_t = 5$ Å and use different major radii from $R = 3r_t$ to $5r_t$, as well as different sink densities. All SIA defects migrated fully in 3D. The results are shown in Fig.\ref{fig:P20100824-1.pdf} as a function of the number density and compared to the theoretical expression for toroidal absorbers, Eq. \eqref{eq:theory_torids}. In Fig. \ref{fig:P20100824-1_Vfrac.pdf}, the OKMC results are plotted as functions of the volume fraction and compared to the theory. It is clear that the theory is not valid for large volume fractions, although for densities of loops typically encountered in materials the theoretical expression is perfectly suitable. Interestingly, both theory and OKMC results collapse onto essentially the same curve, revealing that the volume fraction is the key variable, irrespective of the ratio between radii. Eq. \eqref{eq:theory_torids} can be indeed rewritten as
\begin{equation}
 k^2_{t,3} = \frac{2f_V \sqrt{1-\left(\frac{r_t}{R}\right)^2}}{r_t^2\ln\left(\frac{8R}{r_t}\right)},
\end{equation}
where the volume fraction, $f_V$, is the dominant parameter, compared to $R$, as $r_t$ is constant.

\subsubsection{The dependence on the major radius}

\begin{figure}
 \centering
  \includegraphics[width=\columnwidth]{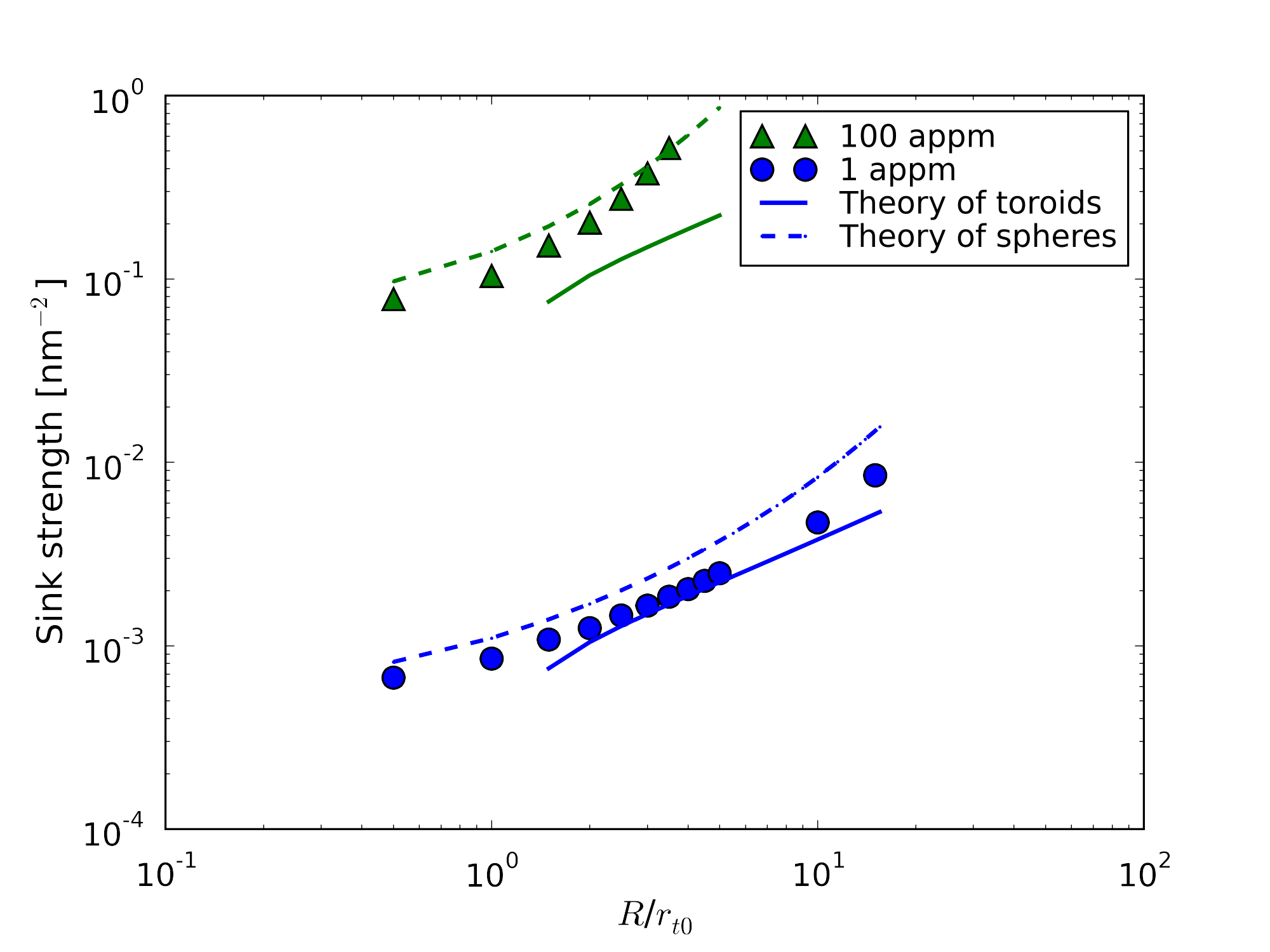}
  \caption{Sink strength of loops for two different number densities (distinguished by different colours), as a function of the ratio between radii when the major one increases: OKMC calculations are compared to theory for both loops and spheres.}
  \label{fig:P20100824-2.pdf}
\end{figure}

\begin{figure}
 \centering
  \includegraphics[width=\columnwidth]{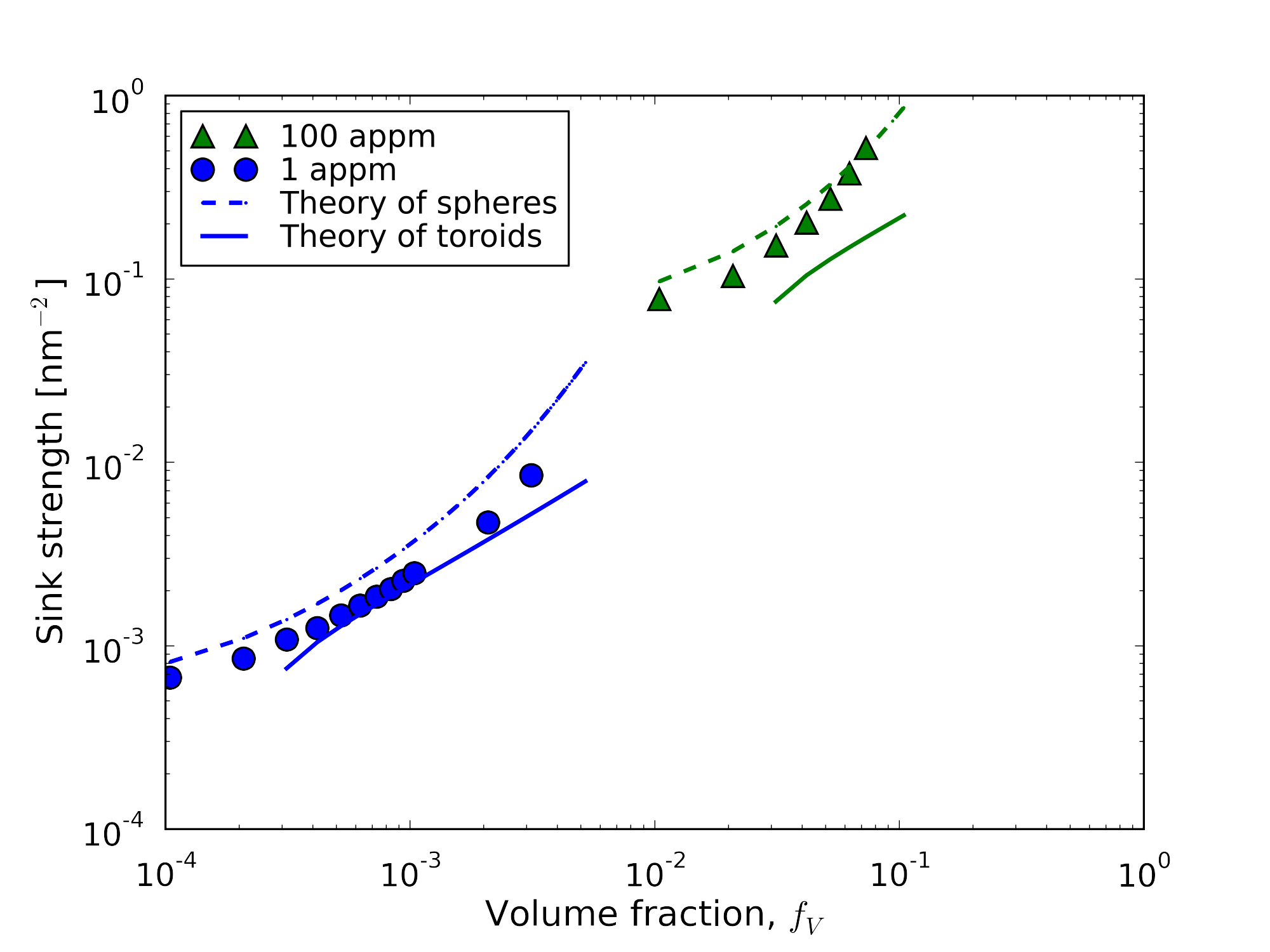}
  \caption{Sink strength of loops for two different number densities (distinguished by different colours)  and different ratios between radii, as a function of the volume fraction: OKMC calculations are compared to theory for both loops and spheres.}
  \label{fig:P20100824-2_Vfrac.pdf}
\end{figure}

We fixed the minor radius at $r_t = 5$ Å and the sink density at $n = 1$ appm ($n=8.46\cdot10^{22}$ m$^{-3}$) or 100 appm ($n=8.46\cdot10^{24}$ m$^{-3}$). The major radii are varied between $R=0.5r_t$ and $R=15r_t$. In Fig \ref{fig:P20100824-2.pdf}, the results are compared to the theory for sink strength of toroids, Eq. \eqref{eq:theory_torids} and spheres, Eq. \eqref{eq:theory_spheres}, where the spherical radius is calculated as $r_s = R+r_t$. In Fig. \ref{fig:P20100824-2_Vfrac.pdf}, the results are plotted versus volume fraction. The theoretical sink strength for toroids differs more at larger volume fractions. The sink strength is better described by the theory for spheres when $R\sim r$.

\subsubsection{The dependence on the minor radius}

\begin{figure}
 \centering
  \includegraphics[width=\columnwidth]{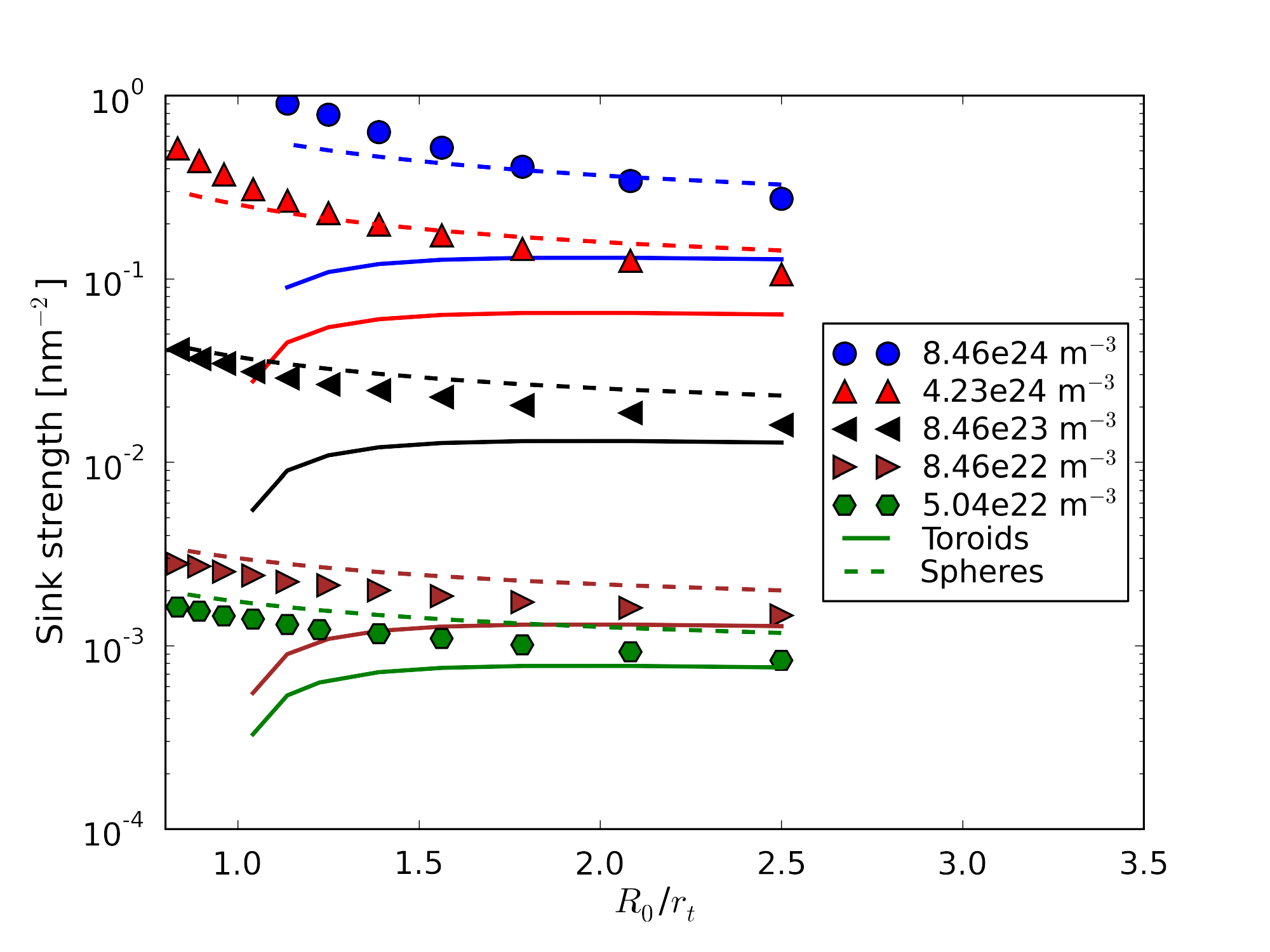}
  \caption{Sink strength of loops for different number densities (distinguished by different colours), as a function of the ratio between radii when the minor one decreases. Dots are OKMC calculations, lines theoretical values for toroidal sinks, dashed lines theoretical values for spherical sinks.}
  \label{fig:P20100824-3.pdf}
\end{figure}

\begin{figure}
 \centering
  \includegraphics[width=\columnwidth]{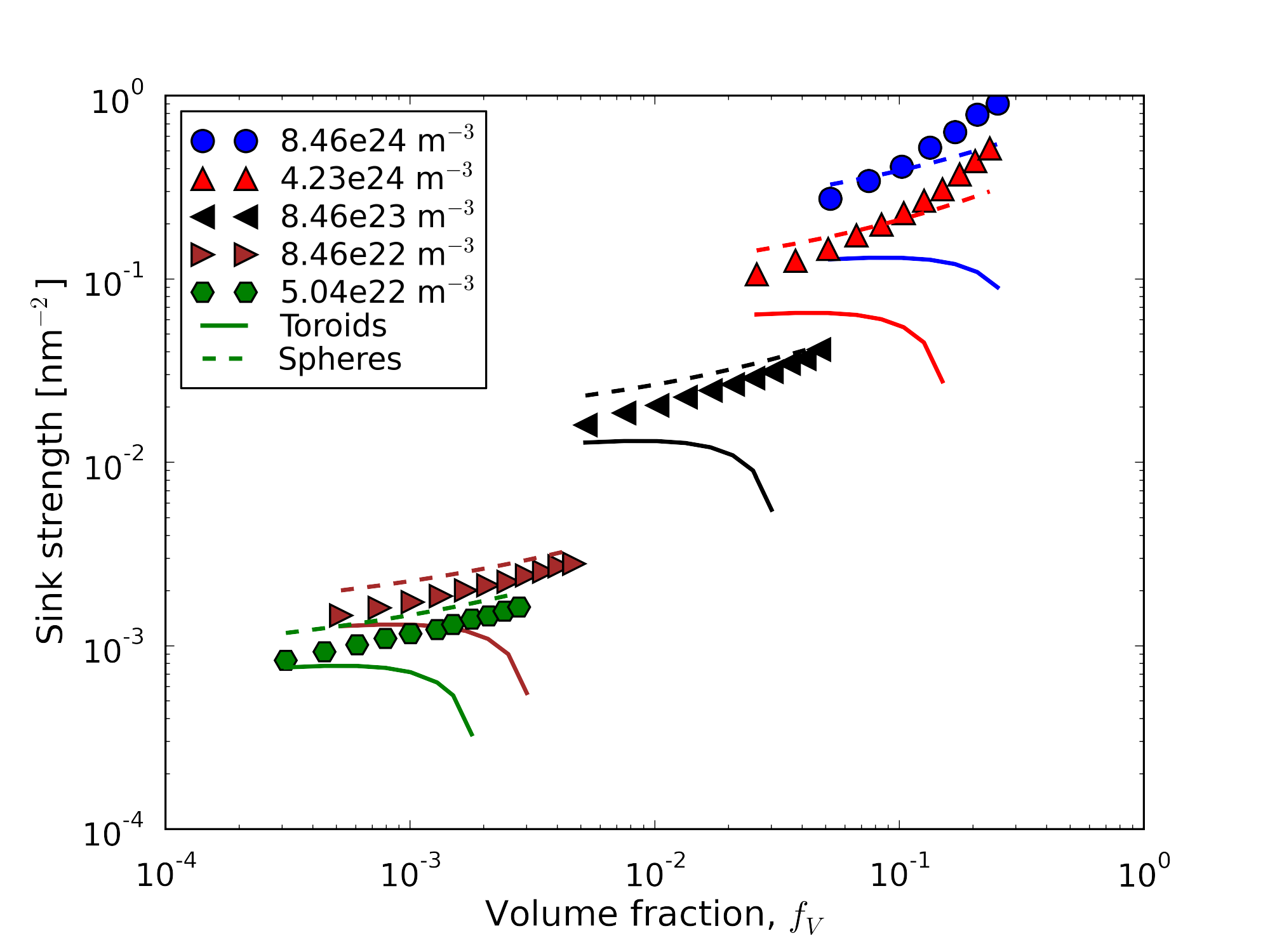}
  \caption{Effect of the minor radius in terms of volume fraction. Dots are our OKMC calculations, lines theoretical values for toroidal sinks and dashed lines the theoretical values for spherical sinks. Different colours correspond to different number densities.}
  \label{fig:P20100824-3_Vfrac.pdf}
\end{figure}

We fixed the major radius at $R=12.5$ Å and the minor radius was varied from $r_t=5$ Å to 13 Å (with $r_t>R$, the toroids become increasingly spherical). The resulting sink strengths are shown for different sink densities in Fig. \ref{fig:P20100824-3.pdf} and in terms of volume fraction in Fig. \ref{fig:P20100824-3_Vfrac.pdf}. The sink strengths are again compared with the theory for toroidal and spherical sinks, Eqs. \eqref{eq:theory_torids} and \eqref{eq:theory_spheres}, respectively. The toroidal theory works best when $r_t$ is small and the spherical theory works best when $r_t$ is large, compared to $R$. The ratio of the radii are here in fact below the constraint of the toroidal theory, which requires $R \gtrsim 3r_t$. Nevertheless, good agreement is reached already at a ratio of 2.5 for the two lowest number densities.

\subsubsection{1D limit for loops}\label{sec:orientation}

We studied the effect on the sink strength of loops when the loops were oriented randomly in four different orientations, compared to having all loops in the same orientation. We used $R=15$ Å, $r_t=5$ Å and $n=10^{24}$ m$^{-3}$. The SIA defects were made to migrate in 1D.

As a result, we got that the sink strength with all sink loops parallel is 0.000279159 nm$^{-2}$ and 0.000310821 nm$^{-2}$ when the loops are randomly oriented; a difference of about 11 \%, smaller than the difference with respect to theoretical expressions. To our knowledge, no theoretical expressions for the sink strength with toroidal sinks and 1D migrating defects exists, so we could not compare results in this case. 

\begin{figure}
 \centering
  \includegraphics[width=\columnwidth]{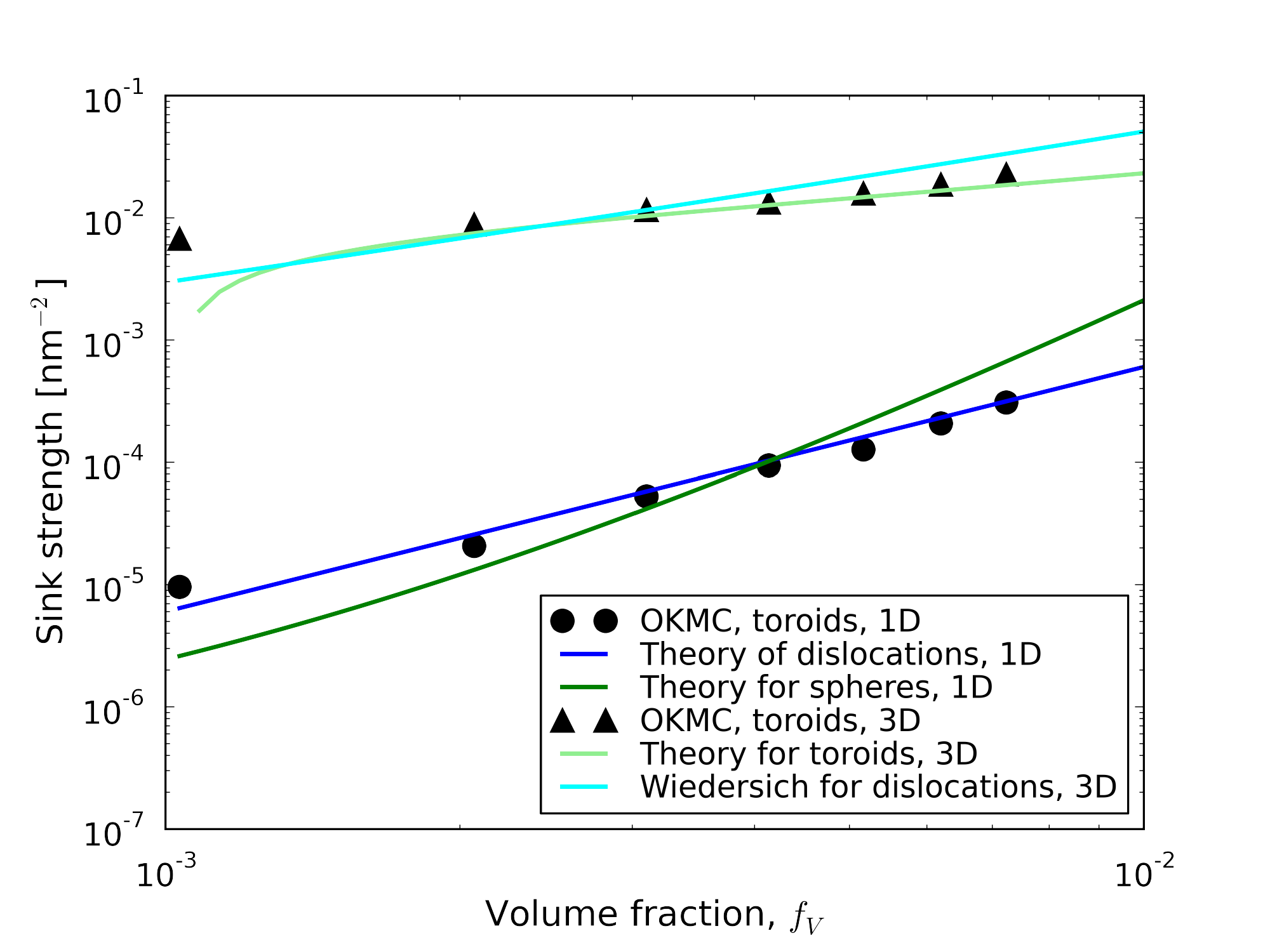}
  \caption{Effect of the major radius in terms of volume fraction for toroidal sinks with 1D migrating defects. $R$ is varied between $0.5r_t$ and $3.5r_t$. The results are compared to the analytical expression for dislocations, using Eq. \eqref{eq:toroids_1d}, and spherical absorbers with 1D migrating defects, Eq. \eqref{eq:theory_spheres_1d}. For comparison, we also include simulation results with 3D migrating defects and the corresponding analytical expression for toroids, Eq. \eqref{eq:theory_torids}, and dislocations (Wiedersich from Table \ref{table:dislocation_eq}).}
  \label{R20121211_1D_Vfrac.pdf}
\end{figure}

Fig. \ref{R20121211_1D_Vfrac.pdf} shows the results of the sink strength of loops for 1D migrating defects as a function of volume fraction for different ratios between radii: $R$ is varied between $0.5r_t$ and $3.5r_t$. Larger $R$ gives larger volume fraction. The loops were randomly oriented. As no analytical expression exist for toroids with 1D migrating defects, we compare the result with the analytical expression for spherical absorbers, Eq. \eqref{eq:theory_spheres_1d}, and with the theory for dislocations with 1D migrating defects. For dislocations, we use Eq. \eqref{eq:barashev} with $\rho^\ast = 1/2\rho_d$, which correspond to random orientation of dislocations \cite{barashev2001reaction} and should be a good approximation for dislocation loops. Using $\rho_d = 2n\pi R$ and $r_d = r_t$, we get from Eq. \eqref{eq:barashev} an approximative expression for the sink strength for loops with 1D migrating defects:
\begin{equation}\label{eq:toroids_1d}
 k_{t,1}^2 = 6(n\pi^2 r_t R)^2
\end{equation}
(The 3D diffusion coefficient is again used; for 1D, divide by 3.) For spherical sinks, we calculate the spherical radius as $r_s = R+r_t$. For comparison, we also include simulation results with 3D migrating defects and the corresponding analytical expression for toroidal sinks, Eq. \eqref{eq:theory_torids}, and dislocations (The Wiedersich equation in Table \ref{table:dislocation_eq}). In the 1D case, we get good agreement with the theory for dislocations and also fair agreement using the theory for spheres. The dislocation comparison does, however, not work as well with 3D migrating defects and we only get fair agreement. The data is in this case, as expected, better described by the theory for toroidal sinks, Eq. \eqref{eq:theory_torids}. The migration regime plays a significantly larger role than the shape of the sinks.

\subsection{The transition from 3D to 1D migration regimes}\label{sec:transition}

The sink strength of an absorber of any shape will depend on the migration regime of the defects it absorbs. Purely 3D migrating defects are more easily absorbed than purely 1D migrating defects and intermediate regimes correspond to intermediate sink strengths, as seen for spherical absorbers in \cite{malerba2007object}. The transition between 1D and 3D migration of the defects has been theoretically investigated by Trinkaus \textit{et al.} in \cite{trinkaus2004reaction}. They proposed a simple master curve which has been shown to agree very well with OKMC simulations of spherical absorbers \cite{malerba2007object}. This master curve depends on two variables $x$ and $y$, defined as
 \begin{align}
  x^2 &= \frac{
    \frac{\delta f^2(\delta) l_{ch}^2 k_3^4}{12k_1^2} + 1}
    {\frac{l_{ch}^2 k_1^4}{12k_1^2} +1} 
    \left( \frac{k_3^2}{k_1^2}\right)
  \left(\frac{k_3^2}{k_1^2}-1\right) ,  \label{eq:dislo_x}\\
  y &= \frac{k^2}{k_1^2},\label{eq:dislo_y}
 \end{align}
where $k_3^2$ is the sink strength in the 3D limit, $k_1^2$ the sink strength in the 1D limit, $k^2$ the sink strength for a given $l_{ch}=d_j\sqrt{n_{ch}}$, the distance travelled in 1D  before change of direction, with $n_{ch}$ being the number of jumps before change of direction. $\delta = D_{tr}/D_{lo}$ is the ratio between transversal and longitudinal diffusion; here  $\delta f^2(\delta) \sim 0$, as transversal motion is not accounted for in our simulations. All sink strengths in the equation refer of course to the same choice of capture radius and dislocation density. The master curve may then be expressed as \cite{trinkaus2004reaction}
\begin{equation}\label{eq:dislo_master}
 y = \frac{1}{2}\left(1+\sqrt{1+4x^2}\right)
\end{equation}

Note that in this definition, the master curve differs slightly from the older and less general definition in \cite{trinkaus20021d}, which was used for the spherical absorbers in \cite{malerba2007object}. That master curve is equivalent to 
\begin{equation}
y'=\frac{1}{2}\left(1+\sqrt{1+\frac{4}{x^2}\left(1-\frac{k_1^2}{k_3^2}\right)}\right)
\end{equation}
with the definition of $x^2$ as in Eq. \eqref{eq:dislo_x}. The difference between how the simulation data satisfy either master curve will only depend on the factor $(1-k_1^2/k_3^2)$, which is essentially equal to 1.

\subsubsection{Dislocations}

\begin{figure}
 \centering
  \includegraphics[width=\columnwidth]{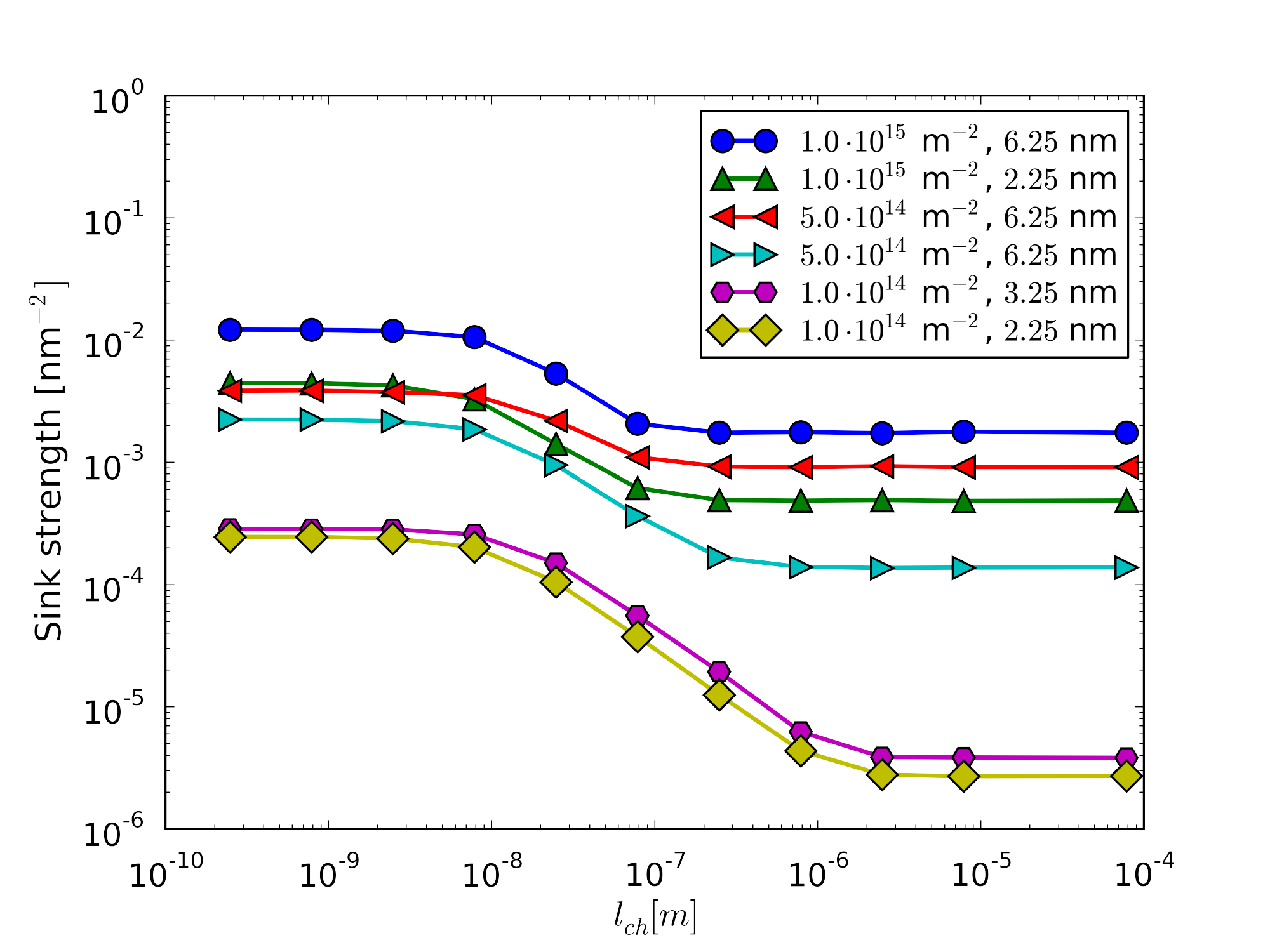}
  \caption{3D to 1D transition as a function of length before change of direction for a few dislocation capture radii (2.25 to 6.25 nm), with a dislocation line density of $10^{15}$ m$^{-2}$}
  \label{6.jpg}
\end{figure}

In Fig. \ref{6.jpg} the 3D to 1D transition of the sink strength of dislocations is shown for a few capture radii (from 2.25 to 6.25 nm), for a dislocation density of $10^{14}$ or $10^{15}$ m$^{-2}$, as a function of the length before change of direction, $l_{ch}$ (more curves could be shown, but they would not add anything qualitatively different). If this figure is compared with Fig. 5 in the published paper on spherical absorbers \cite{malerba2007object}, it can be clearly seen that the transition from 3D to 1D regime seems to occur, in the case of the dislocation, in a much more abrupt way: for a length before change of direction of $\sim$250 nm the 1D regime is already reached, whereas in the case of spherical absorbers a length of at least one order of magnitude larger was needed.

\begin{figure}
 \centering
  \includegraphics[width=\columnwidth]{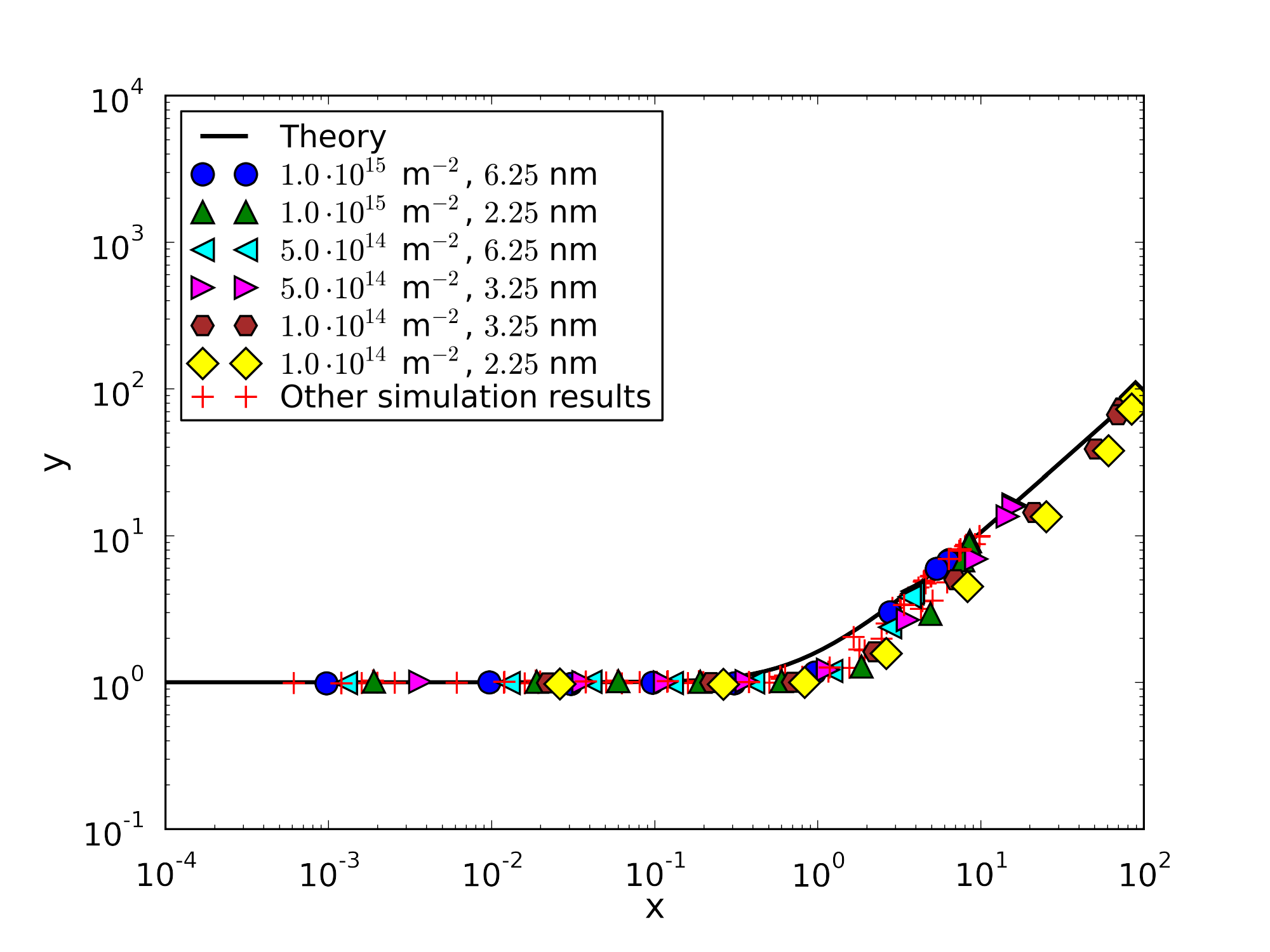}
  \caption{3D to 1D transition in master curve representation for dislocation lines: simulation data points elaborated according to Eqs. \eqref{eq:dislo_x} and \eqref{eq:dislo_y} and the master curve as given in \eqref{eq:dislo_master}. Different dislocation densities [m$^{-2}$] and $r_d$ [nm] have been used. Only representative cases are labelled.}
  \label{7.jpg}
\end{figure}

This more abrupt transition is reflected in the corresponding master curve representation, given in Fig. \ref{7.jpg} for all the conditions (capture radii and dislocation densities) hitherto studied. In this figure, the simulation results were used to calculate $x$ and $y$ as defined by Eqs. \eqref{eq:dislo_x} and \eqref{eq:dislo_y}, respectively. 
It can be seen that, as much as we can say based on the data points we have collected, the master curve expressed as in Eq. \eqref{eq:dislo_master} is reproduced by the data points only in the 1D and 3D limit regions, with, in addition, a tendency to underestimate $y$ in the 3D region, while, again, the 3D to 1D transition is more abrupt than the theoretical master curve predicts.

\subsubsection{Loops}

\begin{figure}
 \centering
  \includegraphics[width=\columnwidth]{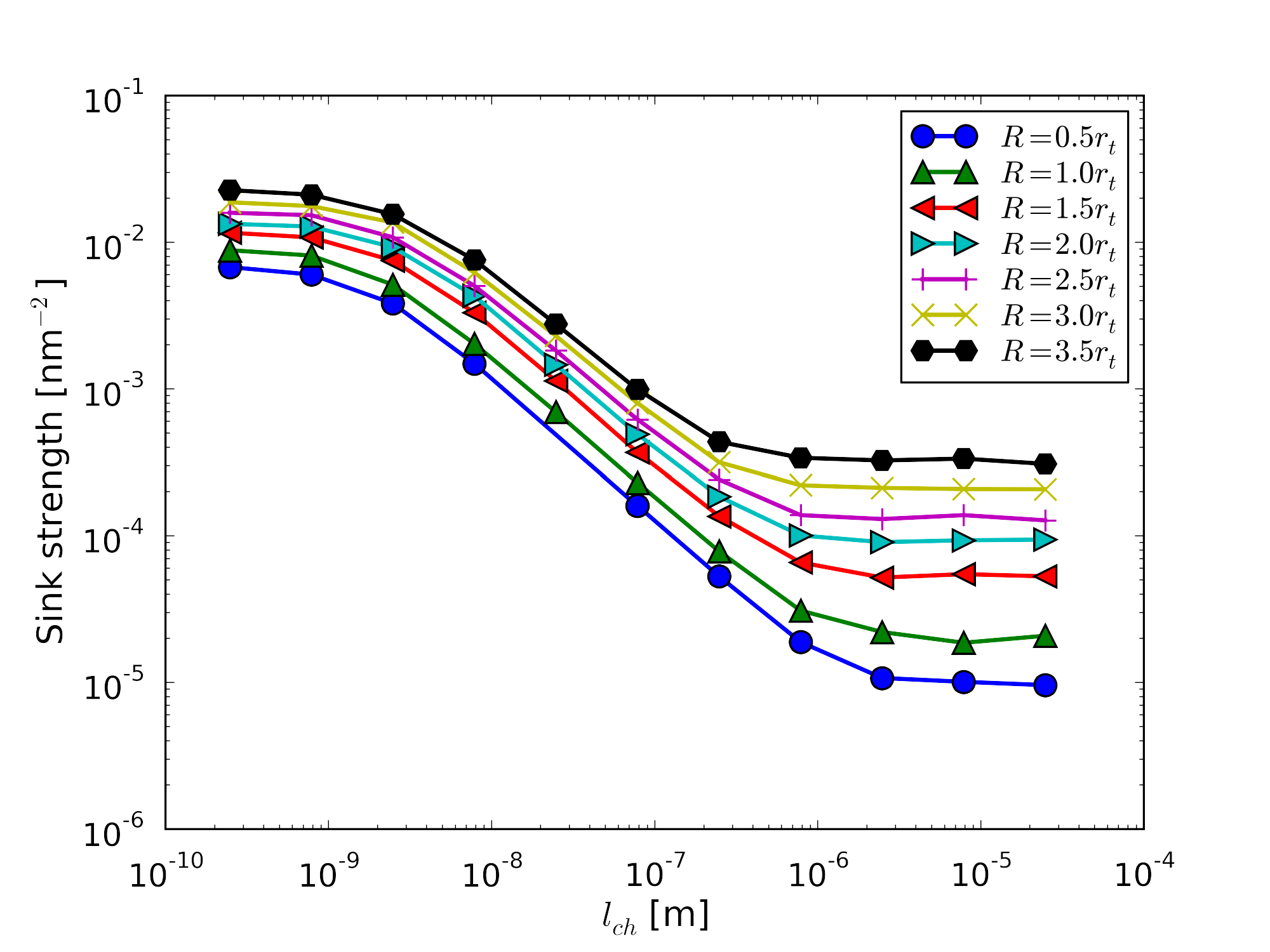}
  \caption{The transition from 3D to 1D migration regimes for different toroidal major radii.}
  \label{fig:R20121211.pdf}
\end{figure}

In Fig. \ref{fig:R20121211.pdf}, the 3D to 1D transition for toroidal absorbers is plotted for different major radii, $R$, \textit{i.e.} different volume fractions, as a function of the distance before change of direction, $l_{ch}$. The sink number density was $n=8.38\cdot10^{23}$ m$^{-3}$. The 1D limit was reached slightly faster for larger $R$ (higher volume fractions), possibly because the loop in this case tends to resemble more and more a dislocation line. With $l_{ch} = 10^{-5}$ m, all cases have reached the 1D limit.

\begin{figure}
 \centering
  \includegraphics[width=\columnwidth]{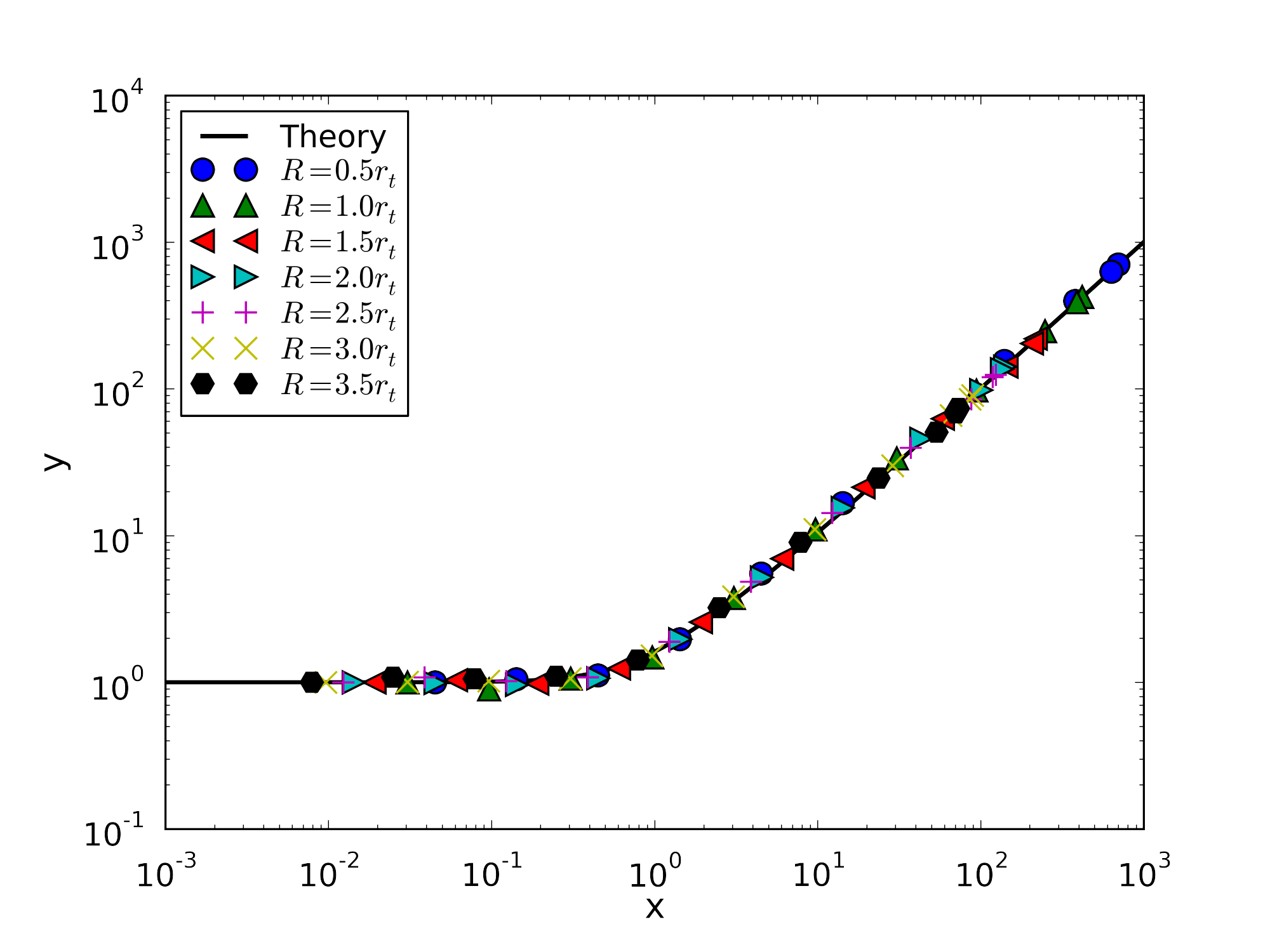}
  \caption{The master curve for different toroidal major radii, compared to the master curve, Eq. \eqref{eq:dislo_master}.}
  \label{fig:R20121211_master.pdf}
\end{figure}

In Fig. \ref{fig:R20121211_master.pdf}, the same data is plotted in the master curve representation and compared to the theory, Eq. \eqref{eq:dislo_master}. The simulation data shows in this case excellent agreement with the theory. The trends are similar to the ones seen for spherical absorbers in \cite{malerba2007object} and thus also less abrupt transition than for dislocations.

\section{Discussion}\label{sec:discussion}

\subsection{Dislocation 1D limit}

\begin{figure}
 \centering
  \includegraphics[width=\columnwidth]{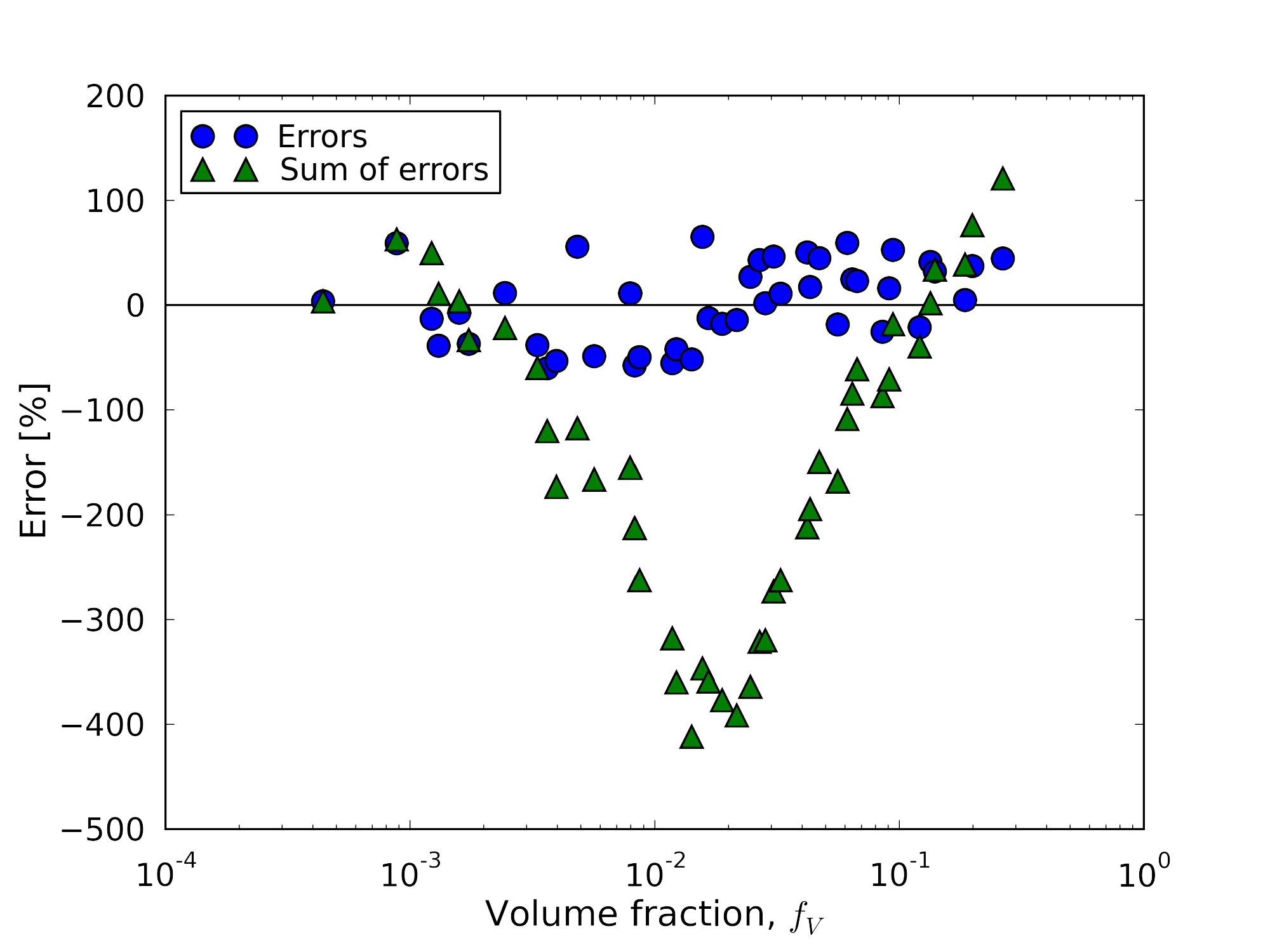}
  \caption{Percentual error committed by taking the simulation result for the sink strength of dislocations in the 1D case, as compared to the reference, chosen to be the theoretical value from Eq. \eqref{eq:barashev}.}
  \label{8.jpg}
\end{figure}

The 45\textdegree{} plot of Fig. \ref{5.jpg} comparing theory and simulation in the 1D limit appears not to be as satisfactory as the same plot for the 3D case (Fig. \ref{3.jpg}), but from there it is difficult to identify any systematic deviation. At first sight, this less good agreement could be attributed to lack of statistics, knowing that it is more difficult to have proper statistics in the 1D case. After closer inspection, however, it appears that the deviation is probably systematic. Namely, for low volume fractions the simulation values tend to be smaller than the theoretical one, while they are larger for high volume fractions. This can be seen in Fig. \ref{8.jpg}, where the percentual error
\begin{equation}
 e(\%) = 100\cdot \frac{k_{simul.}^2-k_{theory}^2}{k_{theory}^2}
\end{equation}
versus sink volume fraction, $f_V$, and the sum of the errors up to the given volume fraction, are plotted. We see that $e(\%)$ is tendentially negative for low $f_V$ and positive for high $f_V$. A sort of critical $f_V$ value, at which the trend is reversed can be identified around 0.02. 

Something similar, though much less spectacular, can be observed, with hindsight, also in the case of the spherical absorbers (see \cite{malerba2007object}). However, in that case the simulation always gave a higher value than the theory and this could be understood in terms of cutting out the tail of very long distances before absorption, given that the time in the simulation is limited and the statistics not very good. This is particularly true for small volume fractions. Since discrepancies were observed also for large volume fractions, however, we might tentatively explain it, in the case of spherical absorbers, with the fact that the simulation may include some degree of order in the distribution of absorbers,  as a consequence of the periodic boundary conditions which is  of course not included in the theoretical expression, that is strictly valid for a random distribution of absorbers. 

In the present case, \textit{i.e.} dislocation lines, however, for low volume fractions the simulation provides smaller sink strength than the theory, so the statistical explanation (elimination of tails of small values) does not hold any more. Indeed in all cases the number of absorbed defects was well in excess of $10^4$, thereby guaranteeing that the values are converged. In addition, in this case we positively know that the simulation is done in an ordered array of dislocations. At the same time, in the theoretical expression this is reflected in the choice of  $\rho^\ast = \rho_d$. We therefore surmise that the theoretical expression might in this case need to be completed with higher order terms that slightly modulate the expression as given in Eq. \eqref{eq:barashev}.

\subsection{Loops}

The sink strengths, calculated for toroidal sinks, are in good agreement with the theoretical expression for toroidal sinks, Eq. \eqref{eq:theory_torids}, when the volume fractions are low enough, $f_V\lesssim10^{-3}$, and the minor ($r_t$) and major ($R$) toroidal radii are not too small or large, compared to each other. Indeed, we get good agreement when $R/r_t>2$ at small volume fractions. When the $r_t$ and $R$ are of comparable sizes ($R/r<2$), the sink strengths are better described by the theory for spherical absorbers, Eq. \eqref{eq:theory_spheres}.

The orientation of the toroidal sinks makes a difference if the defects are migrating in 1D. With 3D migrating defects, the orientations play an insignificant role. Experimentally, we can expect the loops to be oriented randomly, perhaps with the exception of mechanically strained materials.

The sink strength for toroidal sinks with 1D migrating defects can be well described at low densities by the theory for dislocations in random orientations, Eq. \eqref{eq:barashev}. The agreement is only fair with 3D migrating defects and it also remains an open question what happens at higher densities.

\subsection{The master curve}

% The 3D to 1D transition simulations are in this work the most computationally expensive, requiring around two months of CPU time to reach the 1D limit for the used box size and sink number density.

The above-discussed discrepancy for dislocations does not influence the master curve representation, since simulation data in it are compared to other simulation data in order to define $x$ and $y$ (Eqs. \eqref{eq:dislo_x} and \eqref{eq:dislo_y}), so for example in the 1D limit the ratio that gives $y$ (Eq. \eqref{eq:dislo_y}) will tend to unity in any case. The master curve coming out of the simulation, however, has a somewhat different shape from the theoretical one, related to the more abrupt 3D to 1D transition than in the spherical absorber case. In this case the fact that the dislocations in the simulation form an ordered array might play a role, as the master curve expression does not include any variable that takes explicitly into account the presence of order; thus the discrepancy might simply be due to this and to the need to compare to a master curve obtained for a regular array of parallel dislocations.

For the toroidal absorbers, the agreement between simulations and theory is excellent and the values clearly reach the 1D limit, as also was the case for dislocations. The transition also follows the master curve perfectly. The toroidal shapes, as compared to spherical shape, thus, do not alter the validity of the master curve.

\section{Conclusions}\label{sec:conclusions}

There is excellent agreement for both dislocations and toroidal absorbers in the 3D limit between sink strengths estimated statistically in an OKMC simulation and sink strengths obtained using the theoretical expressions available from the literature.

There is, however, only fair agreement for dislocations in the 1D limit between simulation and theory, the reasons of the discrepancy being not the lack of statistics or the inadequacy of the simulation box size. The actual origin of the discrepancy is not established.

The sink strength of toroids approaches the theoretical prediction for spherical absorbers when $r_t \rightarrow R$. Good agreement with theory is reached when the minor radius $r_t$ is small, the major radius between $2r_t$ and $15r_t$, and the volume fraction is low. With 1D migrating defects, the sink strength of toroids is well described at low densities by theory if the loops are considered as dislocations with random orientations.

The master curve is reproduced with excellent agreement using toroidal absorbers and to a good extent also correctly reproduced by the simulation data for the regular array of dislocations. However, the transition between 3D and 1D regime is clearly faster than theoretically predicted in the case of an array of parallel dislocations.

Overall, in any case, we see that theory and OKMC simulations are in mutual agreement for both dislocations and toroidal shaped sinks, such as SIA loops, provided that the volume fractions are small. The master curve representation is globally confirmed by OKMC simulations independently of the shape of the absorbers. 
\section*{Acknowledgement}

This work was carried out as part of the PERFORM60 project of the 7th Euratom
Framework Programme, partially supported by the European Commission, Grant
agreement number FP7-232612. Discussions with S. Golubov and A. Barashev were of fundamental importance to interpret the results of the present work.


%% References


\bibliographystyle{model1a-num-names}
\bibliography{/home/phys-data/people/vjansson/Beam/Articles/vjansson.bib,/home/phys-data/people/vjansson/Beam/Articles/vjansson_publications.bib}






% \end{thebibliography}

\end{document}